# Cooperative Target Realization in Multi-Agent Systems Allowing Choice-Based Actions

Ge Guo[b], *Member, IEEE,* Wing Shing Wong[a*], *Fellow, IEEE,* Zhongchang Liu[a]

*Abstract* — In this paper, we study cooperative multi-agent systems in which the target objective and the controls exercised by the agents are dependent on the choices they made at initial system time. Such systems have been investigated in several recently published papers, mainly from the perspective of system analysis on issues such as control communication complexity, control energy cost and the feasibility of realization of target functions. This paper continues this line of research by developing optimal control design methodology for linear systems that are collaboratively manipulated by multiple agents based on their distributed choices. For *target matrices* that satisfy particular structural constraints, we derive control algorithms that can achieve the specified targets with minimum control cost. We compare state-feedback as well as open-loop control strategies for target realization and extend the optimality result to an arbitrary target matrix. The optimal control solutions are obtained by minimizing the average control cost subject to the set of specified target-state constraints by means of modern variation theory and the Lagrange multiplier method.

## I. Introduction

Research on multi-agent systems has been well motivated by engineering, biological, social and economical problems [22]. For illustration, one can name deep space observation via distributed satellites, automated logistic and manufacturing, teams of robots deployed in hazardous environments [20], home and office building environment control automation, intelligent transportation systems [3], and many others. One important characteristic enabled by using a team of cooperative agents is labor division for handling challenging missions. Multi-agent cooperative control, known under the context of formation flying [13],

Manuscript received Feb 14, 2012. This work was supported by the National Natural Science Foundation of China under grant number 60974013, 61174060, and 60804012; Fok Ying Tong Education Foundation under grant 111066. The second author also acknowledges support from The Chinese University of Hong Kong Shun Hing Institute of Advanced Engineering.
a. Department of Information Engineering, the Chinese University of Hong Kong, Hong Kong, China.
b. School of Control Science and Engineering, Dalian University of Technology, Dalian, 116023, China.
*Corresponding author, email: wswong@ie.cuhk.edu.hk

vehicle platooning [15], flocking [16], rendezvous [10], consensus[18], or swarming [12], have been investigated extensively in recent years under different frameworks, from a host of viewpoints, and for various objectives (e.g. [4], [9], [20] and [21]). The majority of relevant literatures deal with individual utility function optimization for which there are two common categories of solution methodologies: those that achieve individual optimization by assuming constant input from the other team members [9], and those based on decomposing individual agent control inputs into local and global components to minimize individual agent cost functions, such as [20] and [21]. In the past decade, team optimization models based on a single common cost function have also resulted in a considerable number of publications in the literature, such as [7], [17] and [19].

In Wong [23], a new class of distributed control problems was introduced under the premise that the desired system output is a function of the individual choices made by distributed agents. In other words, the multiple agents collaboratively apply controls based on their individual choices to achieve a team target. In environments under limited communication bandwidth constraints, control and information are intricately tied together to form an information-based control system. For systems with two-agents, the minimum amount of information exchange to achieve the desired system output target can be captured by the communication complexity (CC) introduced by Yao [27], resulting in the concept of control communication complexity (CCC). This germ of idea was further expanded in [25] which applied the concept of control communication complexity to an interesting class of nonlinear control systems, known as the Heisenberg-Brockett-Integrator. The authors showed that for this class of systems any target states given in the form of a finite dimensional matrix can be realized using sinusoidal inputs that form closed curves in a product control input space. Moreover, the question of finding distributed controls with minimum control cost was analyzed. In [2], these investigations were further connected with the standard parts optimal control problem ([5] and [6]). The Heisenberg-Brockett-Integrator is a special case of the class of bilinear input-output mapping dynamical systems. In [26], the problem of distributed realization of a target matrix was extended to bilinear input-output mapping systems. These research works lead to new perspectives on distributed control and at the same time raise numerous challenging questions. One of the themes highlighted in [25] and [26] touches on the intricate relationship between the concept of communication complexity and control cost. Roughly speaking, if the agents can jointly achieve a choice-based target at a certain control cost level, it may be possible for them to lower the cost by signaling information of their choices to each other via the dynamic system. Starting from one end of the strategy spectrum, it is possible to design control algorithms that totally ignore information from other agents; namely, one can resort to open-loop controls. On the other hand, one can also consider schemes that allow control decisions to be partially based on information obtained through signaling or state-feedback mechanism.

In this paper we further these investigations by focusing on choice-based actions defined on linear dynamical systems. We also restrict attention to control strategies that do not provide for choice information signaling. The latter assumption significantly simplifies the solution complexity. Yet, the systems considered here are interesting enough of their own rights, since their simplicity is a strong advantage in practical applications. Moreover, these systems serve as a reference point for the more complicated strategies to be investigated in the future.

The main objectives of the paper are twofold: to gain better understanding of the aforementioned systems and to establish optimal control design methodologies. In particular, we derive optimal target-realizing control laws that minimize a common type of control cost function. The optimization problem is solved in the framework of modern variation theory and the Lagrange method. The primary choice-based solutions are given in a basic open-loop form, which is susceptible to external disturbances. To make the result adaptable to systems with state disturbances, we reformulate the primary solutions into feedback control form. These results can be applied to realize target matrices that are compatible, i.e., target matrices whose entries satisfy certain constraints to be explained later. For incompatible target matrices, i.e., target matrices that are not strictly realizable, one can add a terminal error penalty to the cost function in order to derive corresponding optimal choice-based control laws.

We should note that there are fundamental differences between our choice-based action systems and those studies within the framework of the more traditional distributed control and dynamic game theory models (see for example the references in [11]). First of all, as pointed out in [26], in the current research framework, there is no direct communication channel between agents, while in the traditional models, communication between agents, if called for, is typically assumed to be achievable independent of the dynamical system. Second, there is no concept of individual utility functions in our model unlike traditional game theoretic based multi-agent systems. As a consequence, the choice-based target-realization formulation addresses with a brand new problem that requires an untraditional solution approach.

This paper is organized into six sections including the Introduction. Section II contains a motivating example and the detailed problem description. In Section III, the primary result on optimal choice-realizing controls is presented for target matrices that satisfy a basic compatibility condition. Some interesting implications are also observed. In Section IV, we reformulate the primary solutions in the feedback form and present our main feedback control result on choice-based target-realization. We further elaborate our discussion with a numerical example. In Section V we investigate the choice-based realization problem for a general target matrix. Concluding remarks are provided in Section VI.

## II. Motivation example and problem formulation

### *A. Motivation Example*

In [26] a rendezvous problem was presented to motivate the choice-based target realization problem. The problem involves two agents, Alice and Bob, who wish to meet or not to meet in their jogging paths depending on their moods. We will further elaborate on this example through a model in which the agents enter a rectangular shaped park at diagonally opposite corners as shown in Fig. 1. Points A and B represent the positions of Alice and Bob respectively.

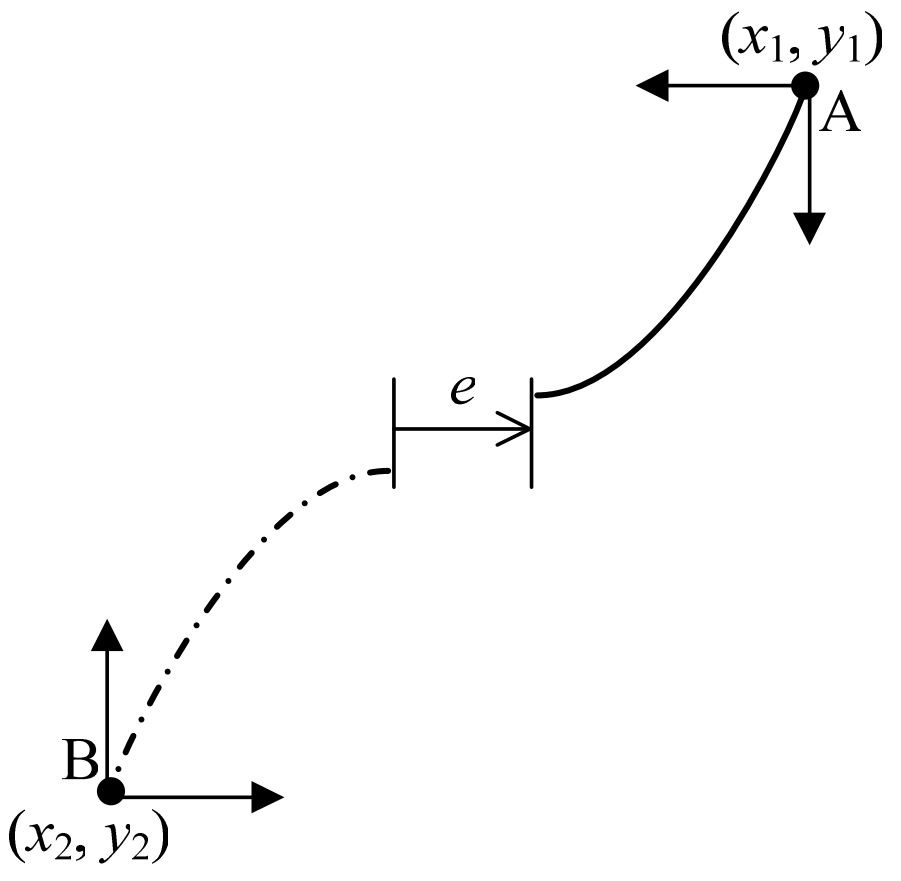

Fig. 1. Path planning for two agents

For simplicity, we assume the vertical velocity components of the points are fixed to be 1 in the direction as indicated in the figure while the horizontal motions are controllable and described by the equations:

$$\begin{bmatrix} \dot{x}_1 \\ \ddot{x}_1 \end{bmatrix} = \begin{bmatrix} 0 & 1 \\ 0 & 0 \end{bmatrix} \begin{bmatrix} x_1 \\ \dot{x}_1 \end{bmatrix} + \begin{bmatrix} 0 \\ 1 \end{bmatrix} u, \tag{1}$$

$$\begin{bmatrix} \dot{x}_2 \\ \ddot{x}_2 \end{bmatrix} = \begin{bmatrix} 0 & 1 \\ 0 & 0 \end{bmatrix} \begin{bmatrix} x_2 \\ \dot{x}_2 \end{bmatrix} + \begin{bmatrix} 0 \\ 1 \end{bmatrix} v, \tag{2}$$

where $u$, $v$ are controls exerted by Alice and Bob respectively. It is assumed that no special communication channels exist between the agents; however, the agents can observe the horizontal coordinate difference:

$$e = x_1 - x_2. \tag{3}$$

At initial time, Alice and Bob each choose from two options with equal probability. Depending on the four equally likely choice pairs, they have a prior agreement on the target system outcome described by a two-by-two matrix with its entries describing one of the two possible event outcomes, namely, the points meet or not meet at a terminal time $T$. The initial choice cannot be changed and are unknown to the other agent. The mathematical definition for the meeting event is simply $e(T) = 0$, (we do not specify the meeting velocities in general, but in this discussion we further specify that $\dot{e}(T) = 0$). On the other hand, the definition for the non-meeting event is less straightforward. In this illustrative example we define it to be $|e(T)| = 5$. Moreover, in order to satisfy the constraint for being realizable we define the target matrix by $\boldsymbol{H} = \begin{bmatrix} 5 & 0 \\ 0 & -5 \end{bmatrix}$. By combining (1) and (2), the system model can be formulated in the form

$$\begin{bmatrix} \dot{e} \\ \ddot{e} \end{bmatrix} = \begin{bmatrix} 0 & 1 \\ 0 & 0 \end{bmatrix} \begin{bmatrix} e \\ \dot{e} \end{bmatrix} + \begin{bmatrix} 0 \\ 1 \end{bmatrix} u + \begin{bmatrix} 0 \\ -1 \end{bmatrix} v + \begin{bmatrix} 0 \\ 1 \end{bmatrix} n(t) . \tag{4}$$

For the stated target matrix, we will show that one can design optimal control laws $\{u_1(t), u_2(t)\}$ for Alice and $\{v_1(t), v_2(t)\}$ for Bob so that if Alice chooses $u_i(t)$ and Bob chooses $v_j(t)$ , the control laws will drive (4) to end with $e(T) = h_{ij}$ , and hence allowing the target matrix be realized as shown in Fig. 2.

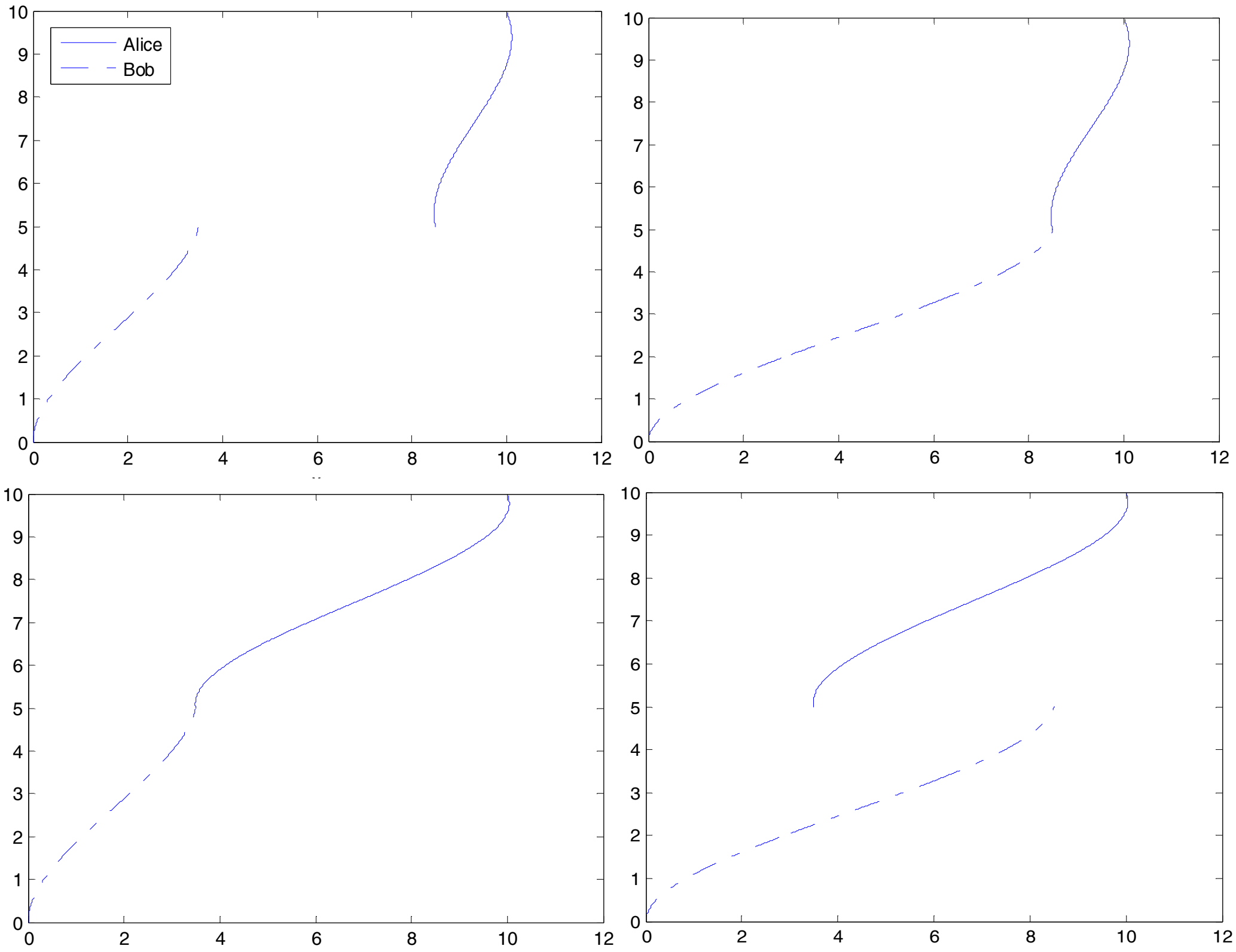

Fig. 2. Trajectories of two particles with initial positions (10, 10) and (0, 0) and target matrix [5 0; 0 -5]

If we change the target matrix to $\boldsymbol{H} = \begin{bmatrix} 5 & 0 \\ 0 & 0 \end{bmatrix}$ it turns out that there are no pre-defined open-loop controls that realize all the targets according to the choices if there is no communication between the agents. In subsequent sections we will analyze this problem and show how control laws with minimal control cost can be designed for a given target matrix.

From this example, it is clear that unlike traditional control problem it is not sufficient to determine just one control function but a set of control laws for the agents to make selections from based on their choices. Moreover, there can be target matrix that cannot be realized in this choice-based action framework if no agent-to-agent communication is allowed.

### B. Problem Setup

Consider a distributed control system with $L$ independent agents

$$\dot{\boldsymbol{x}}(t) = \boldsymbol{A}\boldsymbol{x}(t) + \sum_{l=1}^{L} \boldsymbol{B}_l \boldsymbol{u}_l(t), \ \boldsymbol{x}(t_0) = \boldsymbol{x}_0, \tag{5}$$

where $\boldsymbol{A} \in R^{n\times n}$, $\boldsymbol{B}_l \in R^{n\times m_l}$, $\boldsymbol{x}(t) \in R^n$ is the state, and $\boldsymbol{u}_l(t) \in R^{m_l}$, the control of agent $l$, is selected from $\{\boldsymbol{u}_l^{i_l}(t), i_l = 1,2,\cdots,N_l\}$ solely according to a uniformly distributed choice, $i_l$, made by the agent. It is assumed that no direct communication channel exists between the agents. The target states at terminal time $T$ are represented by an $N_1 \times N_2 \times \cdots \times N_L$ dimensional matrix, or an $L$-order tensor $\boldsymbol{H}$. When agent $l$, for $l = 1,2,\ldots,L,$ chooses $i_l$, our objective is to derive control functions, $\boldsymbol{u}_l^{i_l}$, that minimize the control cost,

$$J = \frac{1}{\prod_{l=1}^{L} N_l} \sum_{i_1=1}^{N_1} \cdots \sum_{i_L=1}^{N_L} \int_{t_0}^{T} (\boldsymbol{u}_l^{i_l}(t))^T \boldsymbol{u}_l^{i_l}(t)\, dt = \int_{t_0}^{T} \sum_{l=1}^{L} \frac{1}{N_l} \sum_{i_l=1}^{N_l} (\boldsymbol{u}_l^{i_l}(t))^T \boldsymbol{u}_l^{i_l}(t)\, dt, \tag{6}$$

which can be interpreted as a quadratic cost averaged over the $N_1 \times N_2 \times \cdots \times N_L$ cases of possible choices, while satisfying the terminal state condition:

$$\boldsymbol{x}(T, \boldsymbol{u}_1^{i_1}, \boldsymbol{u}_2^{i_2}, \cdots, \boldsymbol{u}_L^{i_L}) = \boldsymbol{H}_{i_1 i_2 \ldots i_L}. \tag{7}$$

Here, $\boldsymbol{H}_{i_1 i_2 \ldots i_L}$, the $(i_1, i_2, \ldots, i_L)$ entry of $\boldsymbol{H}$, is the target state when agent $l$ selects the choice $i_l$, for $l = 1,2,\ldots,L$. As explained previously, these choices could be understood as uniformly distributed inputs which are known only to the individual agent at time $t_0$.

**Remark 1**. The problem is challenging for the following reasons: 1) There are multiple agents who make independent choices which are unknown to other agents. 2) The terminal states are given by an $L$-order tensor whose components are dependent on the choices made by all the agents. 3) Standard optimization approach may yield a large number of equations with two-point boundary-values [23] complicated by many constraint conditions. This should become clear when one represents all the terminal states $\boldsymbol{H}_{i_1 i_2 \ldots i_L}$, for $i_l = 1,2,\ldots,N_l, l = 1,2,\ldots,L,$ by the standard formula

$$\boldsymbol{H}_{i_1 i_2 \cdots i_L} = e^{\boldsymbol{A}(T-t_0)}\boldsymbol{x}_0 + \int_{t_0}^{T} e^{\boldsymbol{A}(T-t)}(\boldsymbol{B}_1\boldsymbol{u}_1^{i_1}(t) + \boldsymbol{B}_2\boldsymbol{u}_2^{i_2}(t) + \cdots + \boldsymbol{B}_L\boldsymbol{u}_L^{i_L}(t))\,dt. \tag{8}$$

From simple algebra it follows that for any integers, $i_l$ and $i_l'$ in the set $\{1,2,\cdots N_l\}$, $i_m$ and $i_m'$ in the set $\{1,2,\cdots N_m\}$, $l$ and $m$ in the set $\{1,2,\ldots,L\}$, the following equation holds

$$\boldsymbol{H}_{i_1 i_2 \cdots i_l \cdots i_m \cdots i_L} - \boldsymbol{H}_{i_1 i_2 \cdots i_l \cdots i_m' \cdots i_L} = \boldsymbol{H}_{i_1 i_2 \cdots i_l' \cdots i_m \cdots i_L} - \boldsymbol{H}_{i_1, i_2 \cdots i_l' \cdots i_m' \cdots i_L}. \tag{9}$$

These constraints have been first observed in [26] for the case $L = 2$.

There is a totality of $N_1 \times N_2 \times \cdots \times N_L$ equality constraints of the form (8), of which only $1 + N_1 + N_2 + \cdots + N_L - L$ of them are independent as shown in the following propositions. This fact significantly reduces the complexity of solving the problem.

**Proposition 1:** Define the terminal state set by

$$\mathcal{H} = \{\boldsymbol{H}_{11\cdots 1}\} \cup \{\boldsymbol{H}_{i_1 1\cdots 1} : i_1 = 2,\ldots,N_1\} \cup \{\boldsymbol{H}_{1 i_2 \cdots 1} : i_2 = 2,\ldots,N_2\} \cup \cdots \cup \{\boldsymbol{H}_{11\cdots i_L} : i_L = 2,\ldots,N_L\}.$$

If all entries of a target matrix $\boldsymbol{H}$ satisfy condition (9), then for all $i_l = 1,2,\ldots,N_l, l = 1,2,\ldots,L,$ the vector $\boldsymbol{H}_{i_1 i_2 \cdots i_L}$ is spanned by elements in $\mathcal{H}$. Moreover,

$$\boldsymbol{H}_{i_1 i_2 \cdots i_L} = \boldsymbol{H}_{i_1 1 \cdots 1} + \boldsymbol{H}_{1 i_2 1 \cdots 1} + \cdots \boldsymbol{H}_{1 \cdots 1 i_L} - (L-1)\boldsymbol{H}_{1\cdots 1}. \tag{10}$$

*Proof:* This can be shown by mathematical induction. It is trivial to see that (10) holds for $\boldsymbol{H}_{11\cdots 1}$ and $\boldsymbol{H}_{1\cdots 1, i_l, 1\cdots 1}$ for all $i_l = 2,\ldots,N_l$. Now, for any $k$, $1 < k < L$, assume that (10) holds for all entries $\boldsymbol{H}_{i_1 i_2 \cdots i_L}$ labeled by indices with $k$ or more of them equal to 1. Consider an arbitrary entry of the target matrix labeled by indices with exactly $k$-1 of them equal to 1, $\boldsymbol{H}_{i_1 i_2 \cdots i_L}$. Given any two integers, $m$ and $n$, $1 \le m < n \le L$ we can represent the tuple of indices $i_1 i_2 \cdots i_L$ as $S_1 i_m S_2 i_n S_3$ where $S_1$ stands for string $i_1 \cdots i_{m-1}$, $S_2$ stands for string $i_{m+1} \cdots i_{n-1}$, and $S_3$ stands for string $i_{n+1} \cdots i_L$. It follows from (9) that

$$\boldsymbol{H}_{S_1 i_m S_2 i_n S_3} = \boldsymbol{H}_{S_1 i_m S_2 1 S_3} + \boldsymbol{H}_{S_1 1 S_2 i_n S_3} - \boldsymbol{H}_{S_1 1 S_2 1 S_3}.$$

If $i_m \neq 1$, $i_n \neq 1$, and the index tuple $S_1 i_m S_2 i_n S_3$ has $k$-1 components equal to 1, then $\boldsymbol{H}_{S_1 i_m S_2 1 S_3}$, $\boldsymbol{H}_{S_1 1 S_2 i_n S_3}$ are entries labeled by indices with exactly $k$ components equal to 1 and the index of $\boldsymbol{H}_{S_1 1 S_2 1 S_3}$ has exactly $k$+1 components equal to 1. Hence, all these entries satisfy (10) by induction assumption. It follows then:

$$\boldsymbol{H}_{i_1 i_2 \cdots i_L} = \boldsymbol{H}_{S_1 i_m S_2 i_n S_3} = \boldsymbol{H}_{S_1 i_m S_2 1 S_3} + \boldsymbol{H}_{S_1 1 S_2 i_n S_3} - \boldsymbol{H}_{S_1 1 S_2 1 S_3}$$

$$= \boldsymbol{H}_{i_1 1 \cdots 1} + \boldsymbol{H}_{1 i_2 1 \cdots 1} + \cdots \boldsymbol{H}_{1 \cdots 1 i_L} - (L-1)\boldsymbol{H}_{1 \cdots 1}.$$

Therefore, all entries of the target matrix satisfy (10). □

**Proposition 2:** Suppose all entries of a target matrix $\boldsymbol{H}$ satisfy the constraints stated in (9). If equation (8) holds when the left-hand-side is an element of $\mathcal{H}$, that is, it is of the form $\boldsymbol{H}_{11\cdots 1}$ or $\boldsymbol{H}_{1\cdots 1, i_l, 1 \cdots 1}$ for all $i_l = 2, ..., N_l$, it holds for all entries of $\boldsymbol{H}$.

*Proof:* According to Proposition 1, a general entry of $\boldsymbol{H}$ satisfies equations (10). The proposition follows by substituting the right-hand-side entries of equation (10) by equation (8). □

A target matrix with all its entries satisfying condition (9) is called a *compatible target matrix*, otherwise it is called an *incompatible target matrix.* Choice-based controls that drive a system to realize compatible targets are called *target-achieving* controls. Generally speaking, an arbitrary target matrix does not meet the structural constraint defined by (9) and thus is not realizable. Instead, one can try to find choice-based controls steering the system to terminal states that minimize an extended cost function which includes an averaged terminal state quadratic error penalty as shown in Section V.

## III. PRIMARY TARGET-ACHIEVING CONTROL LAWS

The main result for target-achieving controls for compatible target matrices is derived under the following controllability assumption.

**Assumption 1.** The system in (5) is controllable by each agent, that is, for all $l = 1,2,\ldots,L,$ $(\boldsymbol{A},\boldsymbol{B}_l)$ is a controllable pair, or equivalently the Grammian matrix $\boldsymbol{W}_l = \int_{t_0}^{T} e^{-At}\boldsymbol{B}_l\boldsymbol{B}_l^T e^{-A^T t}dt$ is invertible.

In the rest of this section we present a two-step optimization approach to the optimal target-achieving control problem described above. In the first step of optimization we derive for each agent the class of admissible control functions that satisfy the necessary conditions for optimality irrespective of choices of the other agents; in the second step, the problem is reduced to an optimization problem over an ($L$-1)-dimensional Euclidean space for which there is a unique critical point. This critical point is then shown to be the global minimum.

**Theorem 1.** Consider an $n$-dimensional system with $L$ agents defined by (5) that satisfies Assumption 1. Let $\bar{l}$ be an integer in the set $\{1,2,\ldots,L\}$. The set of *target-achieving* optimal controls minimizing the average control cost (6) subject to (7) is defined as follows: For $i_l = 1,2,\ldots,N_l, l = 1,2,\ldots,L,$

$$\hat{\boldsymbol{u}}_l^{i_l}(t) = \begin{cases} \boldsymbol{B}_l^T e^{-A^T t}\left[\boldsymbol{W}_l^{-1}e^{-AT}(\boldsymbol{H}_{1\cdots i_l\cdots 1} - \boldsymbol{H}_{11\cdots 1}) + \hat{\boldsymbol{P}}_l^1\right], & l \neq \bar{l} \\ \boldsymbol{B}_l^T e^{-A^T t}\left[\boldsymbol{W}_l^{-1}(e^{-AT}\boldsymbol{H}_{1\cdots i_l\cdots 1} - e^{-At_0}\boldsymbol{x}_0 - \sum_{k=1,k\neq\bar{l}}^{L}\boldsymbol{W}_k\hat{\boldsymbol{P}}_k^1)\right], & l = \bar{l} \end{cases}. \tag{11}$$

$$\hat{\boldsymbol{P}}_{\bar{l}}^1 = \boldsymbol{W}_{\bar{l}}^{-1}(e^{-AT}\boldsymbol{H}_{11\cdots 1} - e^{-At_0}\boldsymbol{x}_0 - \sum_{k=1,k\neq\bar{l}}^{L}\boldsymbol{W}_k\hat{\boldsymbol{P}}_k^1),$$

$$\begin{bmatrix} \hat{\boldsymbol{P}}_1^1 \\ \vdots \\ \hat{\boldsymbol{P}}_{\bar{l}-1}^1 \\ \hat{\boldsymbol{P}}_{\bar{l}+1}^1 \\ \vdots \\ \hat{\boldsymbol{P}}_L^1 \end{bmatrix} = \boldsymbol{\Omega}^{-1}\boldsymbol{\Theta}, \tag{12}$$

with

$$\boldsymbol{\Omega}=\begin{bmatrix} \boldsymbol{I}+\boldsymbol{W}_{\bar{l}}^{-1}\boldsymbol{W}_1 & \cdots & \boldsymbol{W}_{\bar{l}}^{-1}\boldsymbol{W}_{\bar{l}-1} & \boldsymbol{W}_{\bar{l}}^{-1}\boldsymbol{W}_{\bar{l}+1} & \cdots & \boldsymbol{W}_{\bar{l}}^{-1}\boldsymbol{W}_L \\ \vdots & \ddots & \vdots & \vdots & \cdots & \vdots \\ \boldsymbol{W}_{\bar{l}}^{-1}\boldsymbol{W}_1 & \cdots & \boldsymbol{I}+\boldsymbol{W}_{\bar{l}}^{-1}\boldsymbol{W}_{\bar{l}-1} & \boldsymbol{W}_{\bar{l}}^{-1}\boldsymbol{W}_{\bar{l}+1} & \cdots & \boldsymbol{W}_{\bar{l}}^{-1}\boldsymbol{W}_L \\ \boldsymbol{W}_{\bar{l}}^{-1}\boldsymbol{W}_1 & & \boldsymbol{W}_{\bar{l}}^{-1}\boldsymbol{W}_{\bar{l}-1} & \boldsymbol{I}+\boldsymbol{W}_{\bar{l}}^{-1}\boldsymbol{W}_{\bar{l}+1} & \cdots & \boldsymbol{W}_{\bar{l}}^{-1}\boldsymbol{W}_L \\ \vdots & \cdots & \vdots & \vdots & \ddots & \vdots \\ \boldsymbol{W}_{\bar{l}}^{-1}\boldsymbol{W}_1 & \cdots & \boldsymbol{W}_{\bar{l}}^{-1}\boldsymbol{W}_{\bar{l}-1} & \boldsymbol{W}_{\bar{l}}^{-1}\boldsymbol{W}_{\bar{l}+1} & \cdots & \boldsymbol{I}+\boldsymbol{W}_{\bar{l}}^{-1}\boldsymbol{W}_L \end{bmatrix},$$

and

$$\boldsymbol{\Theta}=\begin{bmatrix} \boldsymbol{W}_1^{-1}e^{-\boldsymbol{A}T}\sum_{i_1=1}^{N_1}\dfrac{\boldsymbol{H}_{11\cdots1}-\boldsymbol{H}_{i_11\cdots1}}{N_1}+\boldsymbol{W}_{\bar{l}}^{-1}\sum_{i_{\bar{l}}=1}^{N_{\bar{l}}}\dfrac{e^{-\boldsymbol{A}T}\boldsymbol{H}_{1\cdots i_{\bar{l}}\cdots1}-e^{-\boldsymbol{A}t_0}\boldsymbol{x}_0}{N_{\bar{l}}} \\ \vdots \\ \boldsymbol{W}_{\bar{l}-1}^{-1}e^{-\boldsymbol{A}T}\sum_{i_{\bar{l}-1}=1}^{N_{\bar{l}-1}}\dfrac{\boldsymbol{H}_{11\cdots1}-\boldsymbol{H}_{1\cdots i_{\bar{l}-1}\cdots1}}{N_{\bar{l}-1}}+\boldsymbol{W}_{\bar{l}}^{-1}\sum_{i_{\bar{l}}=1}^{N_{\bar{l}}}\dfrac{e^{-\boldsymbol{A}T}\boldsymbol{H}_{1\cdots i_{\bar{l}}\cdots1}-e^{-\boldsymbol{A}t_0}\boldsymbol{x}_0}{N_{\bar{l}}} \\ \boldsymbol{W}_{\bar{l}+1}^{-1}e^{-\boldsymbol{A}T}\sum_{i_{\bar{l}+1}=1}^{N_{\bar{l}+1}}\dfrac{\boldsymbol{H}_{11\cdots1}-\boldsymbol{H}_{1\cdots i_{\bar{l}+1}\cdots1}}{N_{\bar{l}+1}}+\boldsymbol{W}_{\bar{l}}^{-1}\sum_{i_{\bar{l}}=1}^{N_{\bar{l}}}\dfrac{e^{-\boldsymbol{A}T}\boldsymbol{H}_{1\cdots i_{\bar{l}}\cdots1}-e^{-\boldsymbol{A}t_0}\boldsymbol{x}_0}{N_{\bar{l}}} \\ \vdots \\ \boldsymbol{W}_L^{-1}e^{-\boldsymbol{A}T}\sum_{i_L=1}^{N_L}\dfrac{\boldsymbol{H}_{11\cdots1}-\boldsymbol{H}_{11\cdots L}}{N_L}+\boldsymbol{W}_{\bar{l}}^{-1}\sum_{i_{\bar{l}}=1}^{N_{\bar{l}}}\dfrac{e^{-\boldsymbol{A}T}\boldsymbol{H}_{1\cdots i_{\bar{l}}\cdots1}-e^{-\boldsymbol{A}t_0}\boldsymbol{x}_0}{N_{\bar{l}}} \end{bmatrix}. \tag{13}$$

*Proof*. To find the controls that minimize cost function *J*, adjoin the target conditions to *J* with real-valued Lagrange multipliers $\boldsymbol{\lambda}_{i_1i_2\cdots i_L}$ and let

$$\bar{J}=\int_{t_0}^{T}\sum_{l=1}^{L}\frac{1}{N_l}\sum_{i_l=1}^{N_l}(\boldsymbol{u}_l^{i_l}(t))^T\boldsymbol{u}_l^{i_l}(t)dt+\sum_{i_1=1}^{N_1}\sum_{i_2=1}^{N_2}\cdots\sum_{i_L=1}^{N_L}\left[\boldsymbol{\lambda}_{i_1i_2\cdots i_L}^T\left(\boldsymbol{H}_{i_1i_2\cdots i_L}-e^{\boldsymbol{A}(T-t_0)}\boldsymbol{x}_0-\int_{t_0}^{T}e^{\boldsymbol{A}(T-t)}\sum_{l=1}^{L}\boldsymbol{B}_l\boldsymbol{u}_l^{i_l}(t)dt\right)\right], \tag{14}$$

or, equivalently

$$\bar{J}=\sum_{i_1=1}^{N_1}\sum_{i_2=1}^{N_2}\cdots\sum_{i_L=1}^{N_L}\boldsymbol{\lambda}_{i_1i_2\cdots i_L}^T\left(\boldsymbol{H}_{i_1i_2\cdots i_L}-e^{\boldsymbol{A}(T-t_0)}\boldsymbol{x}_0\right)+\int_{t_0}^{T}\Gamma(\boldsymbol{u}_1^{i_1},\boldsymbol{u}_2^{i_2},\cdots,\boldsymbol{u}_L^{i_L},t)dt,$$

where

$$\Gamma(\boldsymbol{u}_1^{i_1},\boldsymbol{u}_2^{i_2},\cdots,\boldsymbol{u}_L^{i_L},t)=\sum_{l=1}^{L}\frac{1}{N_l}\sum_{i_l=1}^{N_l}(\boldsymbol{u}_l^{i_l}(t))^T\boldsymbol{u}_l^{i_l}(t)-\sum_{i_1=1}^{N_1}\sum_{i_2=1}^{N_2}\cdots\sum_{i_L=1}^{N_L}\left[\boldsymbol{\lambda}_{i_1i_2\cdots i_L}^T e^{\boldsymbol{A}(T-t)}\boldsymbol{B}_l\boldsymbol{u}_l^{i_l}(t)\right]. \tag{15}$$

Now we can use standard Lagrange method and the fundamental approach to calculus of variations to find the solution to this optimization problem. Consider the variation in $\bar{J}$ due to variations in the control vectors,

$$\delta\bar{J} = \sum_{l=1}^{L}\left[\int_{t_0}^{T}\left(\frac{2}{N_l}(\boldsymbol{u}_l^{i_l}(t))^T - \sum_{i_1=1}^{N_1}\cdots\sum_{i_{l-1}=1}^{N_{l-1}}\sum_{i_{l+1}=1}^{N_{l+1}}\cdots\sum_{i_L=1}^{N_L}\boldsymbol{\lambda}_{i_1i_2\cdots i_L}^T e^{A(T-t)}\boldsymbol{B}_l\right)\delta\boldsymbol{u}_l^{i_l}(t)dt\right].$$

Here we set $\delta\boldsymbol{x}_0 = 0$ and $\delta\boldsymbol{H}_{i_1i_2\cdots i_L} = 0$, for $i_l = 1,2,\ldots,N_l, l = 1,2,\ldots,L,$ since the initial state and the terminal states are specified. For an extremum, $\delta\bar{J}$ must be zero for arbitrary $\delta\boldsymbol{u}_l^{i_l}(t), l = 1,2,\ldots,L;$ this can happen only if the stationary solutions to $\boldsymbol{u}_l^{i_l}(t)$'s are of the following type

$$\boldsymbol{u}_l^{i_l}(t) = \boldsymbol{B}_l^T e^{-A^T t}\boldsymbol{P}_l^{i_l}, \tag{16}$$

where

$$\boldsymbol{P}_l^{i_l} = \frac{N_l}{2}e^{A^T T}\sum_{i_1=1}^{N_1}\cdots\sum_{i_{l-1}=1}^{N_{l-1}}\sum_{i_{l+1}=1}^{N_{l+1}}\cdots\sum_{i_L=1}^{N_L}\boldsymbol{\lambda}_{i_1i_2\cdots i_L}.$$

In addition to the necessary condition arising from zero first-order variation of $\bar{J}$, the non-negativity of the second-order variation for all values of $\delta\boldsymbol{u}_l^{i_l}(t)$ is easy to verify. Therefore, among all possible types of controls the class of controls given in (16) yields the least average control cost $\bar{J}$. The only thing left is to determine the Lagrange multipliers $\boldsymbol{\lambda}_{i_1i_2\cdots i_L}$, or equivalently the set of values for $\boldsymbol{P}_l^{i_l}$, that minimize the cost function (6). Note that the cost function can now be reformulated as:

$$J = \sum_{l=1}^{L}\frac{1}{N_l}\sum_{i_l=1}^{N_l}(\boldsymbol{P}_l^{i_l})^T\boldsymbol{W}_l\boldsymbol{P}_l^{i_l}. \tag{17}$$

For this reduced problem, the equality constraint in (8) for all $i_l = 1,2,\ldots,N_l, l = 1,2,\ldots,L,$ assumes the form

$$\boldsymbol{H}_{i_1i_2\cdots i_L} = e^{AT}\left(e^{-At_0}\boldsymbol{x}_0 + \sum_{l=1}^{L}\boldsymbol{W}_l\boldsymbol{P}_l^{i_l}\right). \tag{18}$$

According to Proposition 2, if equation (18) holds for all elements in $\mathcal{H}$, it holds for an arbitrary $\boldsymbol{H}_{i_1i_2\cdots i_L}$. As a result, it is sufficient to summarize (18) by the following set of equalities:

$$
\begin{cases}
e^{-AT}\boldsymbol{H}_{11\cdots1}-e^{-At_0}\boldsymbol{x}_0=\sum_{l=1}^{L}\boldsymbol{W}_l\boldsymbol{P}_l^1,\\
e^{-AT}\boldsymbol{H}_{i_11\cdots1}-e^{-At_0}\boldsymbol{x}_0=\boldsymbol{W}_1\boldsymbol{P}_1^{i_1}+\sum_{l=2}^{L}\boldsymbol{W}_l\boldsymbol{P}_l^1, i_1=2,\ldots,N_1,\\
e^{-AT}\boldsymbol{H}_{1i_2\cdots1}-e^{-At_0}\boldsymbol{x}_0=\boldsymbol{W}_2\boldsymbol{P}_2^{i_2}+\sum_{l=1,l\neq2}^{L}\boldsymbol{W}_l\boldsymbol{P}_l^1, i_2=2,\ldots,N_2,\\
\cdots\\
e^{-AT}\boldsymbol{H}_{11\cdots i_L}-e^{-At_0}\boldsymbol{x}_0=\boldsymbol{W}_L\boldsymbol{P}_L^{i_L}+\sum_{l=1}^{L-1}\boldsymbol{W}_l\boldsymbol{P}_l^1, i_L=2,\ldots,N_L.
\end{cases}
\tag{19}
$$

Note that there are $N_1+\cdots N_L+1-L$ equations with $N_1+\cdots+N_L$ variables. This implies there are only $L-1$ free variables. Selecting an arbitrary $\bar{l}\in\{1,2,\ldots,L\}$, define the set of free variables by $\{\boldsymbol{P}_1^1,\cdots,\boldsymbol{P}_{\bar{l}-1}^1,\boldsymbol{P}_{\bar{l}+1}^1,\cdots,\boldsymbol{P}_L^1\}$, and note that:

$$
\boldsymbol{P}_{\bar{l}}^1=\boldsymbol{W}_{\bar{l}}^{-1}(e^{-AT}\boldsymbol{H}_{11\cdots1}-e^{-At_0}\boldsymbol{x}_0-\sum_{k=1,k\neq\bar{l}}^{L}\boldsymbol{W}_k\boldsymbol{P}_k^1)\,. \tag{20}
$$

Based on this and (19), we obtain the following equations for all $\boldsymbol{P}_l^{i_l}$,

$$
\boldsymbol{P}_l^{i_l}=\begin{cases}
\boldsymbol{W}_l^{-1}e^{-AT}(\boldsymbol{H}_{1\cdots i_l\cdots1}-\boldsymbol{H}_{11\cdots1})+\boldsymbol{P}_l^1, & l\neq\bar{l}\\
\boldsymbol{W}_{\bar{l}}^{-1}(e^{-AT}\boldsymbol{H}_{1\cdots i_l\cdots1}-e^{-At_0}\boldsymbol{x}_0-\sum_{k=1,k\neq\bar{l}}^{L}\boldsymbol{W}_k\boldsymbol{P}_k^1), & l=\bar{l}
\end{cases}. \tag{21}
$$

Now, combining (16) and (21) with the cost function (17) leads to an equivalent problem of finding $\overline{\boldsymbol{P}}^1=[(\boldsymbol{P}_1^1)^T\cdots(\boldsymbol{P}_{\bar{l}-1}^1)^T,(\boldsymbol{P}_{\bar{l}+1}^1)^T\cdots(\boldsymbol{P}_L^1)^T]^T$, such that the following function is minimized

$$
\begin{aligned}
J(\overline{\boldsymbol{P}}^1)&=J(\boldsymbol{P}_1^1,\cdots,\boldsymbol{P}_{\bar{l}-1}^1,\boldsymbol{P}_{\bar{l}+1}^1,\cdots,\boldsymbol{P}_L^1)\\
&=\frac{1}{N_{\bar{l}}}\sum_{i_{\bar{l}}=1}^{N_{\bar{l}}}\left[\boldsymbol{W}_{\bar{l}}^{-1}(e^{-AT}\boldsymbol{H}_{1\cdots i_{\bar{l}}\cdots1}-e^{-At_0}\boldsymbol{x}_0-\sum_{k=1,k\neq\bar{l}}^{L}\boldsymbol{W}_k\boldsymbol{P}_k^1)\right]^T\boldsymbol{W}_{\bar{l}}\left[\boldsymbol{W}_{\bar{l}}^{-1}(e^{-AT}\boldsymbol{H}_{1\cdots i_{\bar{l}}\cdots1}-e^{-At_0}\boldsymbol{x}_0-\sum_{k=1,k\neq\bar{l}}^{L}\boldsymbol{W}_k\boldsymbol{P}_k^1)\right],\\
&\quad+\sum_{l=1,l\neq\bar{l}}^{L}\frac{1}{N_l}\sum_{i_l=1}^{N_l}\left[\boldsymbol{W}_l^{-1}e^{-AT}(\boldsymbol{H}_{1\cdots i_l\cdots1}-\boldsymbol{H}_{11\cdots1})+\boldsymbol{P}_l^1\right]^T\boldsymbol{W}_l\left[\boldsymbol{W}_l^{-1}e^{-AT}(\boldsymbol{H}_{1\cdots i_l\cdots1}-\boldsymbol{H}_{11\cdots1})+\boldsymbol{P}_l^1\right].
\end{aligned}
\tag{22}
$$

By means of the equations $\partial J / \partial \boldsymbol{P}_l^1 = 0$ for all $l \neq \bar{l}$, one can show that a critical point for this problem must be given by $\boldsymbol{P}_l^1 = \hat{\boldsymbol{P}}_l^1$, for all $l \neq \bar{l}$, as defined in (12). Moreover, this solution is uniquely determined by the system parameters. We next prove that (12) defines the global minimum of (22). Denoting the minimum of all the eigenvalues of all the $\boldsymbol{W}_l$'s by $\alpha$, which is positive since all $\boldsymbol{W}_l$'s are positive definite, we can have

$$J(\bar{\boldsymbol{P}}^1) = J(\boldsymbol{P}_1^1, \cdots, \boldsymbol{P}_{\bar{l}-1}^1, \boldsymbol{P}_{\bar{l}+1}^1, \cdots, \boldsymbol{P}_L^1)$$

$$\geq \sum_{l=1, l\neq \bar{l}}^{L} \frac{1}{N_l} \sum_{i_l=1}^{N_l} \alpha \left\| \boldsymbol{W}_l^{-1} e^{-AT} (\boldsymbol{H}_{1\cdots i_l \cdots 1} - \boldsymbol{H}_{11\cdots 1}) + \boldsymbol{P}_l^1 \right\|^2$$

$$= \alpha \sum_{l=1, l\neq \bar{l}}^{L} \frac{1}{N_l} \sum_{i_l=1}^{N_l} \left\| \boldsymbol{D}_{i_l} + \boldsymbol{P}_l^1 \right\|^2 \geq \alpha \sum_{l=1, l\neq \bar{l}}^{L} \frac{1}{N_l^2} \left[ \sum_{i_l=1}^{N_l} \left\| \boldsymbol{D}_{i_l} + \boldsymbol{P}_l^1 \right\| \right]^2 \geq \alpha \sum_{l=1, l\neq \bar{l}}^{L} \frac{1}{N_l^2} \left[ \sum_{i_l=1}^{N_l} \left( \left\| \boldsymbol{P}_l^1 \right\| - \left\| \boldsymbol{D}_{i_l} \right\| \right) \right]^2$$

$$\geq \alpha \sum_{l=1, l\neq \bar{l}}^{L} \frac{1}{N_l^2} \left[ N_l \left( \left\| \boldsymbol{P}_l^1 \right\| - \left\| \boldsymbol{D}_l \right\| \right) \right]^2 = \alpha \sum_{l=1, l\neq \bar{l}}^{L} \left( \left\| \boldsymbol{P}_l^1 \right\| - \left\| \boldsymbol{D}_l \right\| \right)^2$$

where $\boldsymbol{D}_{i_l} = \boldsymbol{W}_l^{-1} e^{-AT} (\boldsymbol{H}_{1\cdots i_l \cdots 1} - \boldsymbol{H}_{11\cdots 1})$ and $\left\| \boldsymbol{D}_l \right\| = \max_{i_l \in \{1,2 \ldots N_l\}} \left\| \boldsymbol{D}_{i_l} \right\|$. For any arbitrary positive $C_p$**,** on the circle $\sum_{l=1, l\neq \bar{l}}^{L} \left\| \boldsymbol{P}_l^1 \right\|^2 = C_P^2$, the inequality $\sum_{l=1, l\neq \bar{l}}^{L} \left\| \boldsymbol{P}_l^1 \right\|^2 \geq \frac{1}{L-1} \left( \sum_{l=1, l\neq \bar{l}}^{L} \left\| \boldsymbol{P}_l^1 \right\| \right)^2$ implies $\sum_{l=1, l\neq \bar{l}}^{L} \left\| \boldsymbol{P}_l^1 \right\| \leq \sqrt{L-1} C_P$. Therefore, we have

$$J(\bar{\boldsymbol{P}}^1) \geq \alpha \left( C_P^2 - 2 \sum_{l=1, l\neq \bar{l}}^{L} \left\| \boldsymbol{P}_l^1 \right\| \left\| \boldsymbol{D}_l \right\| + D^2 \right) \geq \alpha \left( C_P^2 - 2d\sqrt{L-1} C_P + D^2 \right)$$

where $D = \sum_{l=1, l\neq \bar{l}}^{L} \left\| \boldsymbol{D}_l \right\|^2$ and $d = \max_{l \in \{1,\ldots,L\}/\{\bar{l}\}} \left\| \boldsymbol{D}_l \right\|$. Since $D$ and $d$ are just constants related to $\boldsymbol{A}$, $\boldsymbol{B}_l$, $\boldsymbol{x}_0$, $\boldsymbol{H}$, and $T$, we can guarantee that $J(\bar{\boldsymbol{P}}^1) > J(\hat{\boldsymbol{P}}_1^1, \cdots, \hat{\boldsymbol{P}}_{\bar{l}-1}^1, \hat{\boldsymbol{P}}_{\bar{l}+1}^1, \cdots, \hat{\boldsymbol{P}}_L^1)$ if $C_P$ is sufficiently large. This indicates that the minimum point cannot lie on the boundary of the bounded domain defined by $\sum_{l=1, l\neq \bar{l}}^{L} \left\| \boldsymbol{P}_l^1 \right\|^2 \leq C_P^2$. Moreover, $\boldsymbol{J}$ is a continuously differentiable function in the region $\sum_{l=1, l\neq \bar{l}}^{L} \left\| \boldsymbol{P}_l^1 \right\|^2 \leq C_P^2$ and must have a global minimum. So the global minimum is an interior point and hence must satisfy the first order necessary condition. As shown before, (12) is the unique critical point derived from the first order necessary condition, therefore it must be the global minimum of (22).

Hence by setting $\boldsymbol{P}_l^1 = \hat{\boldsymbol{P}}_l^1$, for all $l \neq \bar{l}$, and define the remaining $\boldsymbol{P}_l^{i_l}$ via equations (20) and (21), and define the controls via equation (16), we obtain the set of optimal solutions as given in Theorem 1. This completes the proof. □

Theorem 1 has many interesting implications. We observe some of them below. We first specify the result of Theorem 1 to a scalar system:

$$\dot{x}(t) = ax(t) + \sum_{l=1}^{L} b_l u_l(t)\,, \tag{23}$$

where $a$, $b_l$, $x(t)$ and $u_l(t) \in R$.

**Corollary 1.** Consider a scalar system (23) with $L$ agents, and let $\bar{l}$ be an integer in the set $\{1,2,\ldots,L\}$. The set of optimal controls is then given by

$$\hat{u}_l^{i_l}(t) = \begin{cases} e^{-at}\left[\dfrac{1}{b_l g} e^{-aT}(H_{1\cdots i_l \cdots 1} - H_{11\cdots 1}) + b_l \hat{p}_l^1\right], & l \neq \bar{l} \\ e^{-at}\left[\dfrac{1}{b_l g}(e^{-aT} H_{1\cdots i_l \cdots 1} - e^{-at_0} x_0) - \dfrac{1}{b_l}\displaystyle\sum_{k=1,k\neq \bar{l}}^{L} b_k^2 \hat{p}_k^1\right], & l = \bar{l} \end{cases}, \tag{24}$$

$$g = \int_{t_0}^{T} e^{-2at}\,dt,$$

$$\hat{p}_{\bar{l}}^1 = \frac{1}{b_{\bar{l}}^2 g}(e^{-aT} H_{11\cdots 1} - e^{-at_0} x_0) - \frac{1}{b_{\bar{l}}^2}\sum_{k=1,k\neq \bar{l}}^{L} b_k^2 \hat{p}_k^1\,,$$

$$\begin{bmatrix} \hat{p}_1^1 \\ \vdots \\ \hat{p}_{\bar{l}-1}^1 \\ \hat{p}_{\bar{l}+1}^1 \\ \vdots \\ \hat{p}_L^1 \end{bmatrix} = \frac{1}{g}\boldsymbol{\Omega}^{-1}\boldsymbol{\Theta}\,, \tag{25}$$

where for $l$=1,2,…,$L$ and $i_l = 1,2,\cdots,N_l$,

$$
\boldsymbol{\Omega}=\begin{bmatrix}
1+\frac{b_1^2}{b_{\bar l}^2} & \cdots & \frac{b_{\bar l-1}^2}{b_{\bar l}^2} & \frac{b_{\bar l+1}^2}{b_{\bar l}^2} & \cdots & \frac{b_L^2}{b_{\bar l}^2}\\
\vdots & \ddots & \vdots & \vdots & \cdots & \vdots\\
\frac{b_1^2}{b_{\bar l}^2} & \cdots & 1+\frac{b_{\bar l-1}^2}{b_{\bar l}^2} & \frac{b_{\bar l+1}^2}{b_{\bar l}^2} & \cdots & \frac{b_L^2}{b_{\bar l}^2}\\
\frac{b_1^2}{b_{\bar l}^2} & \cdots & \frac{b_{\bar l-1}^2}{b_{\bar l}^2} & 1+\frac{b_{\bar l+1}^2}{b_{\bar l}^2} & \cdots & \frac{b_L^2}{b_{\bar l}^2}\\
\vdots & \vdots & \vdots & \vdots & \ddots & \vdots\\
\frac{b_1^2}{b_{\bar l}^2} & \cdots & \frac{b_{\bar l-1}^2}{b_{\bar l}^2} & \frac{b_{\bar l+1}^2}{b_{\bar l}^2} & \cdots & 1+\frac{b_L^2}{b_{\bar l}^2}
\end{bmatrix},
$$

$$
\boldsymbol{\Theta}=\begin{bmatrix}
\frac{1}{b_1^2}e^{-AT}\sum_{i_1=1}^{N_1}\frac{H_{11\cdots 1}-H_{i_1 1\cdots 1}}{N_1}+\frac{1}{b_{\bar l}^2}\sum_{i_{\bar l}=1}^{N_{\bar l}}\frac{e^{-aT}H_{1\cdots i_{\bar l}\cdots 1}-e^{-at_0}x_0}{N_{\bar l}}\\
\vdots\\
\frac{1}{b_{\bar l-1}^2}e^{-AT}\sum_{i_{\bar l-1}=1}^{N_{\bar l-1}}\frac{H_{11\cdots 1}-H_{1\cdots i_{\bar l-1}\cdots 1}}{N_{\bar l-1}}+\frac{1}{b_{\bar l}^2}\sum_{i_{\bar l}=1}^{N_{\bar l}}\frac{e^{-aT}H_{1\cdots i_{\bar l}\cdots 1}-e^{-at_0}x_0}{N_{\bar l}}\\
\frac{1}{b_{\bar l+1}^2}e^{-aT}\sum_{i_{\bar l+1}=1}^{N_{\bar l+1}}\frac{H_{11\cdots 1}-H_{1\cdots i_{\bar l+1}\cdots 1}}{N_{\bar l+1}}+\frac{1}{b_{\bar l}^2}\sum_{i_{\bar l}=1}^{N_{\bar l}}\frac{e^{-aT}H_{1\cdots i_{\bar l}\cdots 1}-e^{-at_0}x_0}{N_{\bar l}}\\
\vdots\\
\frac{1}{b_L^2}e^{-aT}\sum_{i_L=1}^{N_L}\frac{H_{11\cdots 1}-H_{11\cdots L}}{N_L}+\frac{1}{b_{\bar l}^2}\sum_{i_{\bar l}=1}^{N_{\bar l}}\frac{e^{-aT}H_{1\cdots i_{\bar l}\cdots 1}-e^{-at_0}x_0}{N_{\bar l}}
\end{bmatrix}. \qquad (26)
$$

**Remark 2**. For the regulatory problem, that is, $H_{ij} = 0$, for all $i$ and $j$, the average control cost obtained under the optimal controls $\hat{u}_l^{i_l}(t)$ ($i_l = 1,2,\ldots,N_l, l = 1,2,\ldots,L$) is

$$J_{\min} = \frac{2ax_0^2}{(1-e^{-2aT})\sum_{l=1}^{L} b_l^2}, \text{ for } a \neq 0, \text{ and}$$

$$J_{\min} = \frac{1}{T\sum_{l=1}^{L} b_l^2} x_0^2, \text{ for } a = 0.$$

Especially, for the special case of $T$=1, $L$=2, $b_l$ =1 ($l$=1, 2), the minimum average control cost becomes

$$J_{\min} = \frac{1}{2}x_0^2, \text{ as } a \to 0,$$

which is consistent with the result obtained from traditional single agent control systems. The only difference is that each agent contributes only half of the energy in performing the task. In other words, they both save half of the energy by cooperating.

**Remark 3.** For the simple two-agent case where $t_0 = 0, a = 0, b_1 = b_2 = 1, N_1 = N_2 = 2,$ namely, the two agents, Alice and Bob, each of them has two equally likely choices, the optimal controls are:

$$\hat{u}_1^1(t) = \frac{1}{4T}\left[2H_{11} + H_{12} - H_{21} - 2x_0\right],\ \hat{u}_1^2(t) = \frac{1}{4T}\left[2H_{22} + H_{21} - H_{12} - 2x_0\right],$$

$$\hat{u}_2^1(t) = \frac{1}{4T}\left[2H_{11} - H_{12} + H_{21} - 2x_0\right],\ \hat{u}_2^2(t) = \frac{1}{4T}\left[2H_{22} + H_{12} - H_{21} - 2x_0\right].$$

The average control cost is given by

$$J_{\min} = \frac{1}{4T}\left((H_{11} - x_0)^2 + (H_{22} - x_0)^2 + \frac{1}{2}(H_{21} - H_{12})^2\right). \tag{27}$$

One can see that the average control cost in (27) is composed of two components: one based on the distances between the given target state and the initial state and the other on the level of asymmetry of the target states in their spatial distribution. Therefore, the cost would be lower if the controller is assigned for a set of terminal states that are closer to the initial state or if the prescribed terminal states are more symmetrically distributed. One could argue that the first criterion is more dominant since the distance separation component consists of two terms both of which have larger weights in comparison with the spatial distribution cost. In fact, we can express the optimal control cost in terms of the distance of the target states from the initial state, since it can be rewritten as

$$J_{\min} = \frac{1}{4T}\left((H_{11} - x_0)^2 + (H_{22} - x_0)^2 + \frac{1}{2}\left[(H_{21} - x_0) - (H_{12} - x_0)\right]^2\right)$$

$$= \frac{1}{4T}\left((H_{11} - x_0)^2 + (H_{22} - x_0)^2 + \frac{1}{2}(H_{12} - x_0)^2 + \frac{1}{2}(H_{21} - x_0)^2 - (H_{21} - x_0)(H_{12} - x_0)\right).$$

**Remark 4.** For the special case in which $L$ = 2, $N_1 = N_2 = 1$, namely, the two agents each has a single choice and hence a unique target state $H$, the problem reduces to its traditional choice-free version for which the optimal controls are obtained, for $a$ = 0, as

$$\hat{u}_1(t) = \frac{b_1}{(b_1^2 + b_2^2)T}(H - x_0),\ \hat{u}_2(t) = \frac{b_2}{(b_1^2 + b_2^2)T}(H - x_0).$$

The average control cost is given by

$$\tilde{J}_{\min} = \frac{(H - x_0)^2}{(b_1^2 + b_2^2)T}.$$

If Alice and Bob are to achieve the four target states $H_{11}$, $H_{12}$, $H_{21}$, and $H_{22}$ one after another in the four single choice cases, the average control cost needed, for $b_1 = b_2 = 1$, is

$$\tilde{J}_{\min} = \frac{1}{8T}[(H_{11} - x_0)^2 + (H_{22} - x_0)^2 + (H_{12} - x_0)^2 + (H_{21} - x_0)^2].$$

It can be shown that, this average control cost is not larger than that obtained in the choice-based action case shown in (27).

## IV. TARGET-ACHIEVING STATE-FEEDBACK CONTROL

The target-achieving controls described in section IV are functions of the initial state and the target states and hence they are essentially open-loop control laws. Since open-loop controller is susceptible to external disturbances, it is preferable to use a feedback configuration for the controller. In this session, we describe a set of target-achieving feedback control laws, which is derived directly from the primary optimal control result presented in the previous section. (See [14] for a corresponding single target problem.)

For the sake of simplicity, we consider systems with two agents here, that is $L$ =2:

$$\dot{\boldsymbol{x}}(t) = \boldsymbol{A}\boldsymbol{x}(t) + \boldsymbol{B}\boldsymbol{u}(t) + \boldsymbol{C}\boldsymbol{v}(t), \tag{28}$$

where $\boldsymbol{A} \in R^{n\times n}$, $\boldsymbol{B} \in R^{n\times m_u}$ and $\boldsymbol{C} \in R^{n\times m_v}$, $\boldsymbol{x}(t) \in R^n$, $\boldsymbol{u}(t) \in R^{m_u}$ and $\boldsymbol{v}(t) \in R^{m_v}$ are the control signals, with $\boldsymbol{u}(t) \in \{\boldsymbol{u}_i(t), i = 1,2,\cdots N_u\}$ and $\boldsymbol{v}(t) \in \{\boldsymbol{v}_j(t), j = 1,2,\cdots N_v\}$. The following control cost and target states are associated with system (28):

$$J=\int_{t_0}^{T}\left(\frac{1}{N_u}\sum_{i=1}^{N_u}\boldsymbol{u}_i^T(t)\boldsymbol{u}_i(t)+\frac{1}{N_v}\sum_{j=1}^{N_v}\boldsymbol{v}_j^T(t)\boldsymbol{v}_j(t)\right)dt\,, \tag{29}$$

$$\boldsymbol{x}[T,\boldsymbol{x}(t_0),\boldsymbol{u}_i,\boldsymbol{v}_j]=\boldsymbol{H}_{ij}\,. \tag{30}$$

According to Theorem 1, the optimal control laws are given by,

$$\hat{\boldsymbol{u}}_i(t)=\boldsymbol{B}^T e^{-A^T t}\Big[\boldsymbol{W}_{\boldsymbol{B}}^{-1}e^{-AT}\boldsymbol{H}_{i1}$$

$$+(\boldsymbol{I}+\boldsymbol{W}_{\boldsymbol{C}}^{-1}\boldsymbol{W}_{\boldsymbol{B}})^{-1}\left(\boldsymbol{W}_{\boldsymbol{C}}^{-1}e^{-AT}\sum_{j=1}^{N_v}\frac{\boldsymbol{H}_{1j}-\boldsymbol{H}_{11}}{N_v}-\boldsymbol{W}_{\boldsymbol{B}}^{-1}e^{-AT}\sum_{k=1}^{N_u}\frac{\boldsymbol{H}_{k1}}{N_u}-\boldsymbol{W}_{\boldsymbol{C}}^{-1}e^{-At_0}\boldsymbol{x}(t_0)\right)\Bigg], \tag{31}$$

$$\hat{\boldsymbol{v}}_j(t)=\boldsymbol{C}^T e^{-A^T t}\Big[\boldsymbol{W}_{\boldsymbol{C}}^{-1}e^{-AT}\boldsymbol{H}_{1j}$$

$$+(\boldsymbol{I}+\boldsymbol{W}_{\boldsymbol{B}}^{-1}\boldsymbol{W}_{\boldsymbol{C}})^{-1}\left(\boldsymbol{W}_{\boldsymbol{B}}^{-1}e^{-AT}\sum_{i=1}^{N_u}\frac{\boldsymbol{H}_{i1}-\boldsymbol{H}_{11}}{N_u}-\boldsymbol{W}_{\boldsymbol{C}}^{-1}e^{-AT}\sum_{k=1}^{N_v}\frac{\boldsymbol{H}_{1k}}{N_v}-\boldsymbol{W}_{\boldsymbol{B}}^{-1}e^{-At_0}\boldsymbol{x}(t_0)\right)\Bigg], \tag{32}$$

where

$$\boldsymbol{W}_{\boldsymbol{B}}=\int_{t_0}^{T}e^{-At}\boldsymbol{B}\boldsymbol{B}^T e^{-A^T t}dt\,,\ \boldsymbol{W}_{\boldsymbol{C}}=\int_{t_0}^{T}e^{-At}\boldsymbol{C}\boldsymbol{C}^T e^{-A^T t}dt\,. \tag{33}$$

Note that the above controls are valid for arbitrary $t_0$ and $\boldsymbol{x}(t_0)$, so they must hold for all $t\in[t_0,T)$ and $\boldsymbol{x}(t)$. Therefore, we can derive the following result.

**Theorem 2**. For the system (28), the target states are realized by the feedback controls given by

$$\hat{\boldsymbol{u}}_i(t)=-\boldsymbol{B}^T\boldsymbol{K}(t)\boldsymbol{x}(t)+\boldsymbol{L}_{\boldsymbol{u}i}(t)\,,$$

$$\hat{\boldsymbol{v}}_j(t)=-\boldsymbol{C}^T\boldsymbol{K}(t)\boldsymbol{x}(t)+\boldsymbol{L}_{\boldsymbol{v}j}(t)\,, \tag{34}$$

for all $i=1,2,\cdots N_u$, $j=1,2,\cdots N_v$, where

$$\boldsymbol{K}(t)=e^{-A^T t}(\overline{\boldsymbol{W}}_{\boldsymbol{B}}+\overline{\boldsymbol{W}}_{\boldsymbol{C}})^{-1}e^{-At}\,, \tag{35}$$

$$\boldsymbol{L}_{\boldsymbol{u}i}(t)=\boldsymbol{B}^T e^{-A^T t}\left[\overline{\boldsymbol{W}}_{\boldsymbol{B}}^{-1}e^{-AT}\boldsymbol{H}_{i1}+(\boldsymbol{I}+\overline{\boldsymbol{W}}_{\boldsymbol{C}}^{-1}\overline{\boldsymbol{W}}_{\boldsymbol{B}})^{-1}\left(\overline{\boldsymbol{W}}_{\boldsymbol{C}}^{-1}e^{-AT}\sum_{j=1}^{N_v}\frac{\boldsymbol{H}_{1j}-\boldsymbol{H}_{11}}{N_v}-\overline{\boldsymbol{W}}_{\boldsymbol{B}}^{-1}e^{-AT}\sum_{k=1}^{N_u}\frac{\boldsymbol{H}_{k1}}{N_u}\right)\right],$$

$$\boldsymbol{L}_{\boldsymbol{v}j}(t)=\boldsymbol{C}^T e^{-A^T t}\left[\overline{\boldsymbol{W}}_{\boldsymbol{C}}^{-1}e^{-AT}\boldsymbol{H}_{1j}+(\boldsymbol{I}+\overline{\boldsymbol{W}}_{\boldsymbol{B}}^{-1}\overline{\boldsymbol{W}}_{\boldsymbol{C}})^{-1}\left(\overline{\boldsymbol{W}}_{\boldsymbol{B}}^{-1}e^{-AT}\sum_{i=1}^{N_u}\frac{\boldsymbol{H}_{i1}-\boldsymbol{H}_{11}}{N_u}-\overline{\boldsymbol{W}}_{\boldsymbol{C}}^{-1}e^{-AT}\sum_{k=1}^{N_v}\frac{\boldsymbol{H}_{1k}}{N_v}\right)\right], \tag{36}$$

and $\overline{\boldsymbol{W}}_{\boldsymbol{B}} = \int_t^T e^{-At}\boldsymbol{B}\boldsymbol{B}^T e^{-A^T t} dt$, $\overline{\boldsymbol{W}}_{\boldsymbol{C}} = \int_t^T e^{-At}\boldsymbol{C}\boldsymbol{C}^T e^{-A^T t} dt$. (37)

**Remark 5.** Theorem 2 appears to be a useful tool of designing feedback control laws for choice-based distributed action systems. But it suffers from a critical shortcoming that does not occur in traditional linear systems with single target. Namely, the foregoing feedback controls have singularities at time *t* = *T*. This point can be easily verified from the derived control laws in the example studied under Remark 3 by replacing *T* with (*T*-*t*) and $x_0$ with *x*(*t*). Thus, the controls in (34)-(36) will blow up as *t* approaches the terminal time *T*, although the total control effect on (28) is bounded owing to a cancellation of the singularities with one another. The cause of this problem is due to the nature of the feedback controls defined by (34) which totally ignore the information conveyed by the state trajectory up to time *t*. Instead, a new open-loop problem is solved for each *t* anew, but with a diminishing amount of time to respond. This leads eventually to the singularity in the control.

To address the singularity problem and test the robustness of the feedback system, we propose a hybrid control strategy for linear systems with state disturbances of the form:

$$\dot{\boldsymbol{x}}(t) = \boldsymbol{A}\boldsymbol{x}(t) + \boldsymbol{B}\boldsymbol{u}(t) + \boldsymbol{C}\boldsymbol{v}(t) + \boldsymbol{n}(t), \tag{38}$$

where $\boldsymbol{n}(t) \in R^n$ is a noise process. The idea is to use the feedback control obtained in Theorem 2 to handle disturbances until some time $T' < T$ and switch to open-loop control given by (31) and (32) after $T'$. The threshold time $T'$ can be determined in principle by solving an optimization problem.

As an example, let us consider the two moving particles introduced in Section II. The target states are given by $[h_{ij}\ 0]^T$, for $i = 1, 2, \cdots N_u$ and $j = 1, 2, \cdots N_v$. Note that (4) is controllable by both agents, so the set of open-loop optimal controls can be solved utilizing Theorem 1 and is given by

$$\hat{u}_{oi}(t) = \left[\frac{-3}{(T-t_0)^2} + \frac{6(t-t_0)}{(T-t_0)^3}\right] e(t_0) + \left[\frac{-2}{T-t_0} + \frac{3(t-t_0)}{(T-t_0)^2}\right] \dot{e}(t_0)$$
$$+ \left[\frac{3}{(T-t_0)^2} - \frac{6(t-t_0)}{(T-t_0)^3}\right] \left[2h_{i1} - h_{11} - \frac{1}{N_u}\sum_{k=1}^{N_u} h_{k1} + \frac{1}{N_v}\sum_{j=1}^{N_v} h_{1j}\right],$$

$$\hat{v}_{oj}(t)=\left[\frac{3}{(T-t_0)^2}-\frac{6(t-t_0)}{(T-t_0)^3}\right]e(t_0)+\left[\frac{2}{T-t_0}-\frac{3(t-t_0)}{(T-t_0)^2}\right]\dot{e}(t_0)$$
$$+\left[\frac{-3}{(T-t_0)^2}+\frac{6(t-t_0)}{(T-t_0)^3}\right]\left[2h_{1j}-h_{11}+\frac{1}{N_u}\sum_{i=1}^{N_u}h_{i1}-\frac{1}{N_v}\sum_{k=1}^{N_v}h_{1k}\right] \tag{39}$$

and the feedback controls according to Theorem 2 are

$$\hat{u}_{fi}(t)=\frac{-3}{(T-t)^2}e(t)-\frac{2}{T-t}\dot{e}(t)+l_{ui}(t),$$

$$\hat{v}_{fj}(t)=\frac{3}{(T-t)^2}e(t)+\frac{2}{T-t}\dot{e}(t)+l_{vj}(t),$$

where

$$l_{ui}(t)=\frac{3}{(T-t)^2}\left[2h_{i1}-h_{11}-\frac{1}{N_u}\sum_{k=1}^{N_u}h_{k1}+\frac{1}{N_v}\sum_{j=1}^{N_v}h_{1j}\right],$$

$$l_{vj}(t)=\frac{-3}{(T-t)^2}\left[2h_{1j}-h_{11}+\frac{1}{N_u}\sum_{i=1}^{N_u}h_{i1}-\frac{1}{N_v}\sum_{k=1}^{N_v}h_{1k}\right].$$

To demonstrate the control result, we performed a simulation study for a two-by-two target matrix:

$$\boldsymbol{H}=\begin{bmatrix}10 & 0\\ 0 & -10\end{bmatrix}.$$

Note that this target matrix is compatible. The starting points of the two particles are (5, 10) and (0, 0) respectively and thus $e(0)=5$. We assume that the terminal time $T = 1$ second, the switch time $T' = 0.6$ second, and $\boldsymbol{n}(t)$ is a band-limited white noise.

The motion trajectories of the two agents produced by the primary open-loop control laws derived in Section III along with the said hybrid control laws are shown in Fig. 3. The horizontal distance variations are depicted in Fig.4. From these figures, it can be seen that with the feedback control laws being utilized in the first 0.6 seconds, the system could reach given target states at terminal time with small deviations. In contrast, the pure open-loop control laws are much sensitive to external disturbances. Last, Fig. 5 indicates that the hybrid controls significantly reduce the control cost.

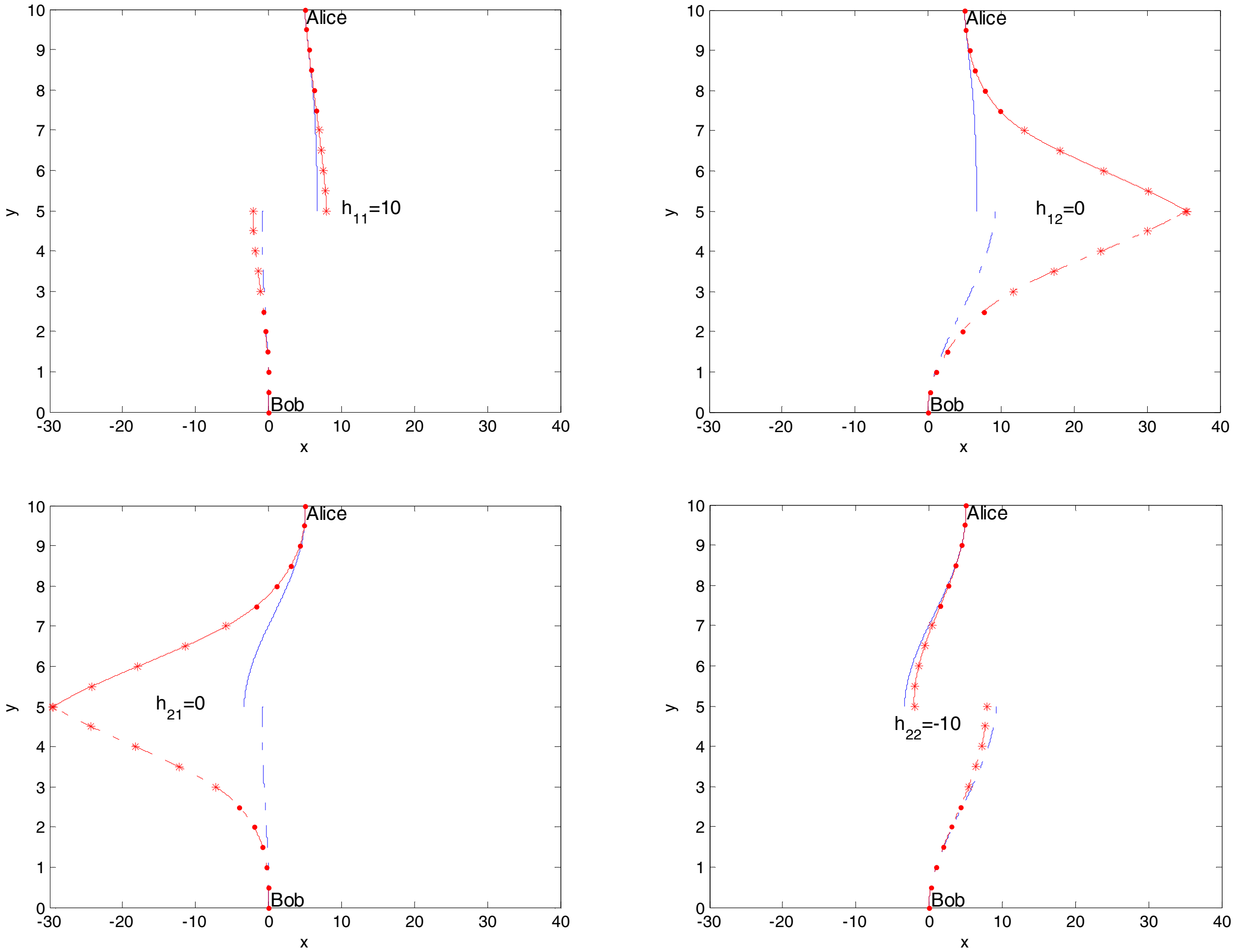

Fig. 3. Motion trajectories of Alice and Bob respectively.

Blue line: open-loop control;

Red point-line: feedback trajectory segment, red star-line: open-loop trajectory segment.

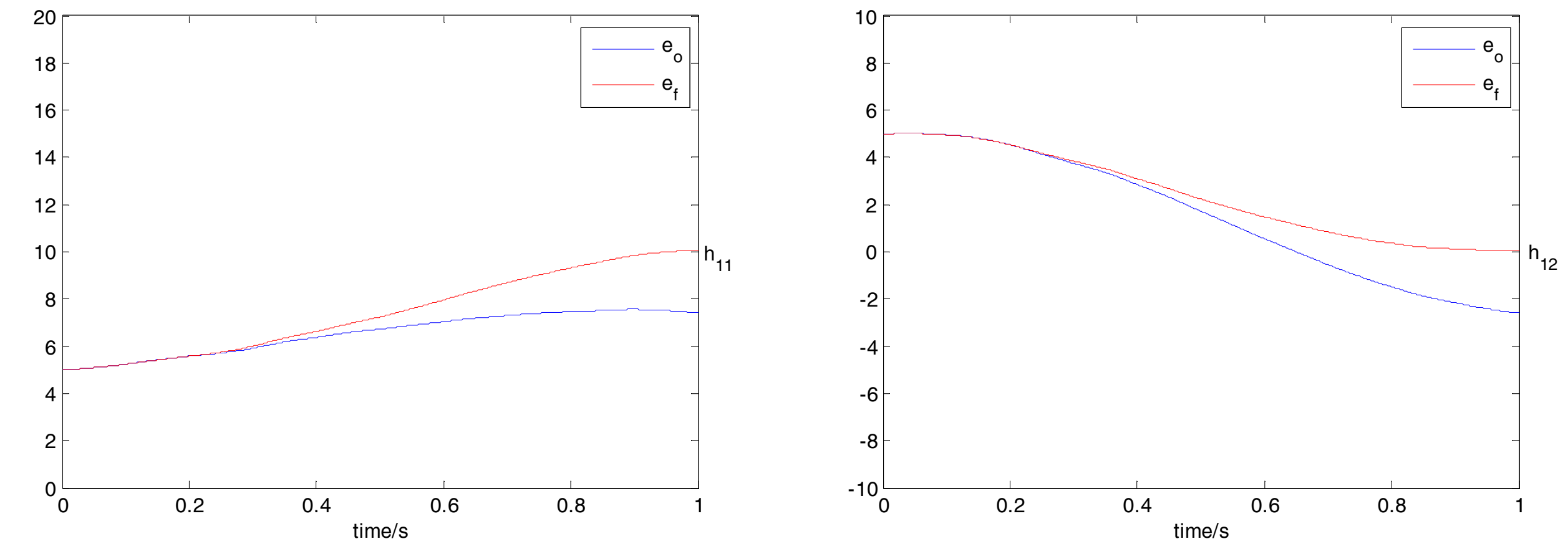

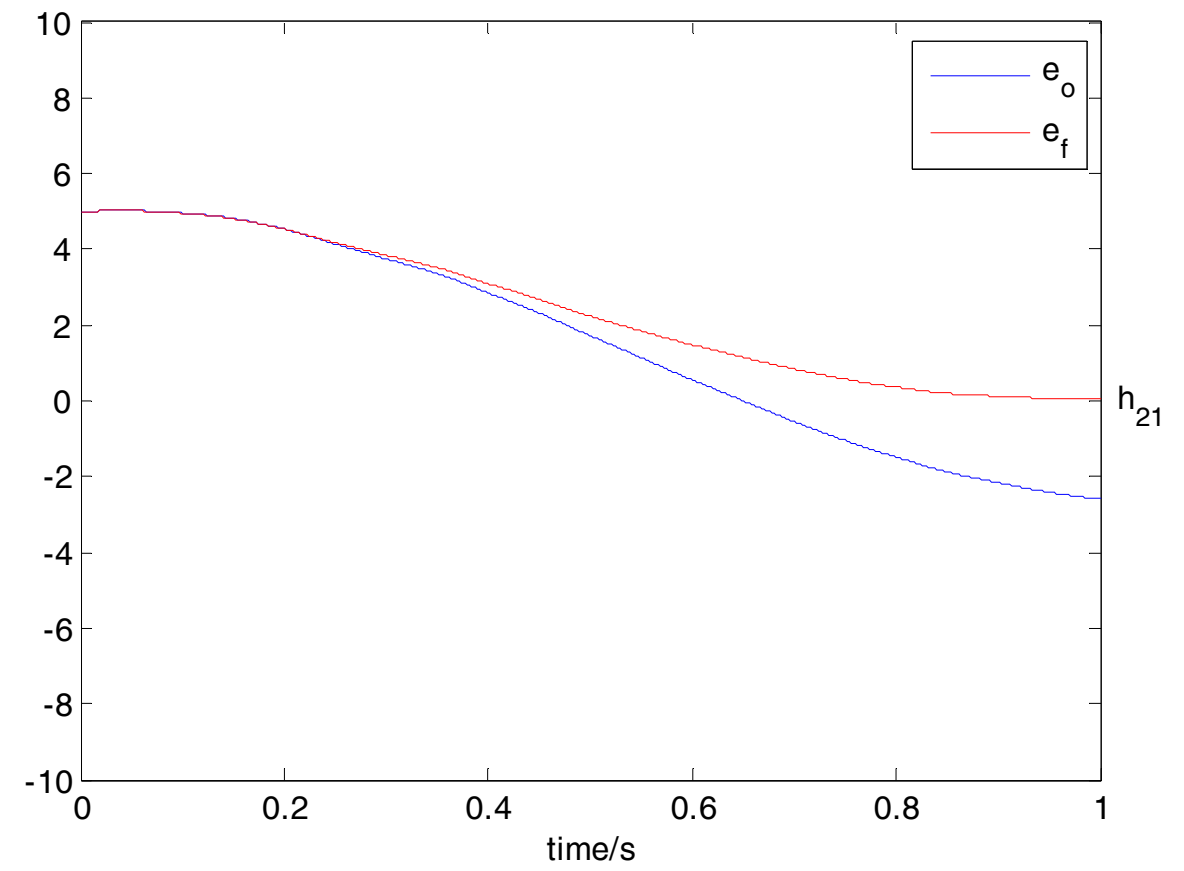

Fig. 4. Trajectories of the horizontal distances:

$e_o$ --pure open-loop control; $e_f$ -- hybrid control.

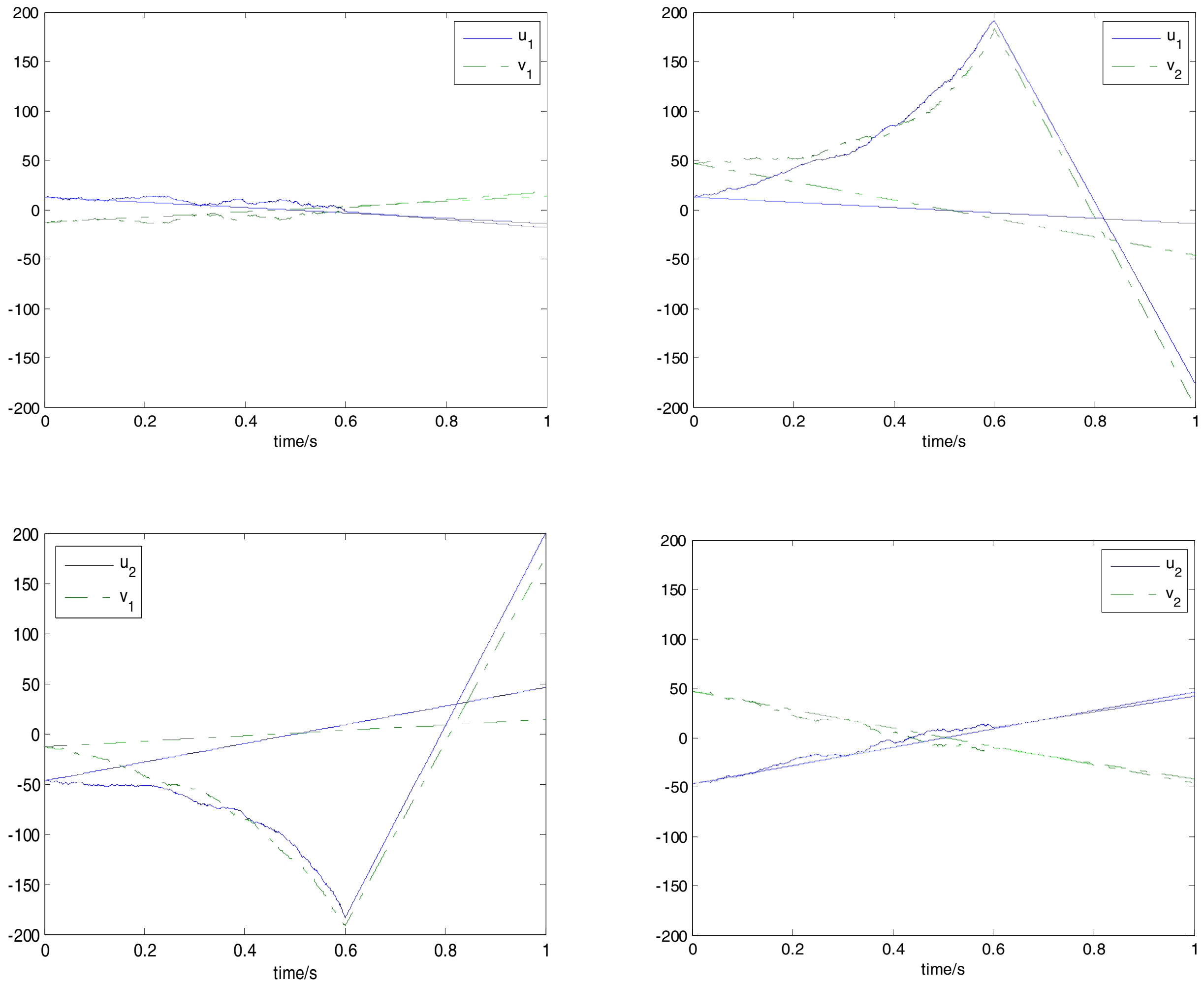

Fig. 5. Quasi-straight lines show pure open-loop control law pairs and curved lines are control pairs with hybrid control.

## V. Target-approaching feedback control

In this section, we aim to find controls for general target matrices, including those that are not target compatible. Specifically, our objective is to derive optimal feedback controls that minimize a weighted sum of the quadratic terminal state error and the control cost, averaged over all possible choices.

Again, consider an $n$-dimensional two-agent system (28). Let $\boldsymbol{x}_{ij}(t)$ be the trajectory at time $t$ when the two agents choose $\boldsymbol{u}_i$ and $\boldsymbol{v}_j$ as their control inputs respectively. Let $\boldsymbol{H}$ be the $N_u$-by-$N_v$ target matrix with arbitrary entries. Our objective in this section is to establish a set of control laws such that all $\boldsymbol{x}_{ij}(T)$ 's are close to $\boldsymbol{H}_{ij}$ on the average. To this end, define the following cost function

$$J = \int_{t_0}^{T} \left( \frac{1}{N_u} \sum_{i=1}^{N_u} \boldsymbol{u}_i^T(t)\boldsymbol{u}_i(t) + \frac{1}{N_v} \sum_{j=1}^{N_v} \boldsymbol{v}_j^T(t)\boldsymbol{v}_j(t) \right) dt + \frac{f}{N_u N_v} \sum_{i=1}^{N_u} \sum_{j=1}^{N_v} \left[\boldsymbol{x}_{ij}(T) - \boldsymbol{H}_{ij}\right]^T \left[\boldsymbol{x}_{ij}(T) - \boldsymbol{H}_{ij}\right] \tag{40}$$

for some positive weighting parameter, $f$.

Using similar argument lines as in the proof of Theorem 1 and 2, we obtained the following result.

**Theorem 3:** For an $n$-dimensional two-agent system in (28), the controls minimizing the cost function (40) are given by

$$\hat{\boldsymbol{u}}_i = -\boldsymbol{K}_{\boldsymbol{u}}(t)\boldsymbol{x}(t) + \boldsymbol{L}_{\boldsymbol{u}i}(t), \tag{41}$$

$$\hat{\boldsymbol{v}}_j = -\boldsymbol{K}_{\boldsymbol{v}}(t)\boldsymbol{x}(t) + \boldsymbol{L}_{\boldsymbol{v}j}(t), \tag{42}$$

where

$$\boldsymbol{K}_{\boldsymbol{u}}(t) = f\boldsymbol{B}^T e^{-A^T t}\left(f\overline{\boldsymbol{W}}_{\boldsymbol{B}} + e^{-(A+A^T)T}\right)^{-1}\left[\boldsymbol{I} - f\overline{\boldsymbol{W}}_{\boldsymbol{C}}\left(e^{-(A+A^T)T} + f(\overline{\boldsymbol{W}}_{\boldsymbol{B}} + \overline{\boldsymbol{W}}_{\boldsymbol{C}})\right)^{-1}\right]e^{-At}, \tag{43}$$

$$\boldsymbol{K}_{\boldsymbol{v}}(t) = f\boldsymbol{C}^T e^{-A^T t}\left(f\overline{\boldsymbol{W}}_{\boldsymbol{C}} + e^{-(A+A^T)T}\right)^{-1}\left[\boldsymbol{I} - f\overline{\boldsymbol{W}}_{\boldsymbol{B}}\left(e^{-(A+A^T)T} + f(\overline{\boldsymbol{W}}_{\boldsymbol{B}} + \overline{\boldsymbol{W}}_{\boldsymbol{C}})\right)^{-1}\right]e^{-At}, \tag{44}$$

$$\boldsymbol{L}_{\boldsymbol{u}i}(t) = -f\boldsymbol{B}^T e^{-A^T t}\left(f\overline{\boldsymbol{W}}_{\boldsymbol{B}} + e^{-(A+A^T)T}\right)^{-1}$$
$$\times\left[\frac{f}{N_u N_v}\overline{\boldsymbol{W}}_{\boldsymbol{C}}\left(e^{-(A+A^T)T} + f(\overline{\boldsymbol{W}}_{\boldsymbol{B}} + \overline{\boldsymbol{W}}_{\boldsymbol{C}})\right)^{-1} e^{-AT}\sum_{k=1}^{N_u}\sum_{j=1}^{N_v}\boldsymbol{H}_{kj} - \frac{e^{-AT}}{N_v}\sum_{j=1}^{N_v}\boldsymbol{H}_{ij}\right], \tag{45}$$

$$\boldsymbol{L}_{vj}(t) = -f\boldsymbol{C}^T e^{-A^T t}\left(f\overline{\boldsymbol{W}}_{\boldsymbol{C}} + e^{-(A+A^T)T}\right)^{-1}$$

$$\times\left[\frac{f}{N_u N_v}\overline{\boldsymbol{W}}_{\boldsymbol{B}}\left(e^{-(A+A^T)T} + f(\overline{\boldsymbol{W}}_{\boldsymbol{B}} + \overline{\boldsymbol{W}}_{\boldsymbol{C}})\right)^{-1} e^{-AT}\sum_{i=1}^{N_u}\sum_{k=1}^{N_v}\boldsymbol{H}_{ik} - \frac{e^{-AT}}{N_u}\sum_{i=1}^{N_u}\boldsymbol{H}_{ij}\right]. \quad (46)$$

*Proof:* Define the following Hamiltonian

$$\Gamma = \frac{1}{N_u}\sum_{i=1}^{N_u}\boldsymbol{u}_i^T(t)\boldsymbol{u}_i(t) + \frac{1}{N_v}\sum_{j=1}^{N_v}\boldsymbol{v}_j^T(t)\boldsymbol{v}_j(t) + \sum_{i=1}^{N_u}\sum_{j=1}^{N_v}\boldsymbol{\lambda}_{ij}^T(\boldsymbol{A}\boldsymbol{x}(t) + \boldsymbol{B}\boldsymbol{u}_i(t) + \boldsymbol{C}\boldsymbol{v}_j(t)), \quad (47)$$

in which $\boldsymbol{\lambda}_{ij} \in R^n$ ($i = 1, 2, \cdots, N_u$, $j = 1, 2, \cdots, N_v$.) According to similar approaches used in the proof of Theorem 1, we obtain the following equalities:

$$\boldsymbol{\lambda}_{ij}(t) = e^{-A^T(t-t_0)}\boldsymbol{\lambda}_{ij}(t_0) \quad (48)$$

and

$$\boldsymbol{u}_i = -\frac{1}{2}N_u\boldsymbol{B}^T e^{-A^T(t-t_0)}\boldsymbol{P}_i, \quad \boldsymbol{v}_j = -\frac{1}{2}N_v\boldsymbol{C}^T e^{-A^T(t-t_0)}\boldsymbol{Q}_j, \quad (49)$$

where

$$\boldsymbol{P}_i = \sum_{j=1}^{N_v}\boldsymbol{\lambda}_{ij}(t_0), \quad \boldsymbol{Q}_j = \sum_{i=1}^{N_u}\boldsymbol{\lambda}_{ij}(t_0). \quad (50)$$

We claim that the controls in (49) minimize the Hamiltonian, since the second order partial derivatives of $\Gamma$ with respect to $\boldsymbol{u}_i$ and $\boldsymbol{v}_j$ are easily verified to be positive-definite. Hence, the remaining task is to calculate explicit expressions for $\boldsymbol{P}_i$ and $\boldsymbol{Q}_j$. To this end, first by combining (49) and system equation (28) we solve the terminal state at $T$ as follows

$$\boldsymbol{x}_{ij}(T) = e^{A(T-t_0)}\boldsymbol{x}(t_0) - \left(\frac{N_u}{2}e^{AT}\boldsymbol{W}_{\boldsymbol{B}}e^{A^T t_0}\boldsymbol{P}_i + \frac{N_v}{2}e^{AT}\boldsymbol{W}_{\boldsymbol{C}}e^{A^T t_0}\boldsymbol{Q}_j\right) \quad (51)$$

with $\boldsymbol{W}_{\boldsymbol{B}}$ and $\boldsymbol{W}_{\boldsymbol{C}}$ defined in (33).

According to the transversality condition (refer to [1]), we have

$$\boldsymbol{\lambda}_{ij}(T) = \frac{\partial}{\partial\boldsymbol{x}_{ij}(T)}\left\{\frac{f}{N_u N_v}\sum_{i=1}^{N_u}\sum_{j=1}^{N_v}\left[\boldsymbol{x}_{ij}(T) - \boldsymbol{H}_{ij}\right]^T\left[\boldsymbol{x}_{ij}(T) - \boldsymbol{H}_{ij}\right]\right\}$$

$$= \frac{2f}{N_u N_v}\left[\boldsymbol{x}_{ij}(T) - \boldsymbol{H}_{ij}\right]. \tag{52}$$

From (48), we know that

$$\boldsymbol{\lambda}_{ij}(T) = e^{-\boldsymbol{A}^T(T-t_0)}\boldsymbol{\lambda}_{ij}(t_0) \tag{53}$$

which when compared with (52) yields

$$\boldsymbol{\lambda}_{ij}(t_0) = \frac{2fe^{\boldsymbol{A}^T(T-t_0)}}{N_u N_v}\left[\boldsymbol{x}_{ij}(T) - \boldsymbol{H}_{ij}\right]. \tag{54}$$

Furthermore,

$$\sum_{i=1}^{N_u}\boldsymbol{P}_i = \sum_{j=1}^{N_v}\boldsymbol{Q}_j = \sum_{i=1}^{N_u}\sum_{j=1}^{N_v}\boldsymbol{\lambda}_{ij}(t_0) = \frac{2f}{N_u N_v}e^{\boldsymbol{A}^T(T-t_0)}\left[\sum_{i=1}^{N_u}\sum_{j=1}^{N_v}\boldsymbol{x}_{ij}(T) - \sum_{i=1}^{N_u}\sum_{j=1}^{N_v}\boldsymbol{H}_{ij}\right]. \tag{55}$$

Combining (51) and (55), we obtain

$$\sum_{i=1}^{N_u}\sum_{j=1}^{N_v}\boldsymbol{x}_{ij}(T) = N_u N_v e^{\boldsymbol{A}(T-t_0)}\boldsymbol{x}(t_0) - fe^{\boldsymbol{A}T}(\boldsymbol{W_B} + \boldsymbol{W_C})e^{\boldsymbol{A}^T T}\left[\sum_{i=1}^{N_u}\sum_{j=1}^{N_v}\boldsymbol{x}_{ij}(T) - \sum_{i=1}^{N_u}\sum_{j=1}^{N_v}\boldsymbol{H}_{ij}\right],$$

from which we solve that

$$\sum_{i=1}^{N_u}\sum_{j=1}^{N_v}\boldsymbol{x}_{ij}(T) = e^{-\boldsymbol{A}^T T}\left[e^{-(\boldsymbol{A}+\boldsymbol{A}^T)T} + f(\boldsymbol{W_B} + \boldsymbol{W}_C)\right]^{-1}\left[N_u N_v e^{-\boldsymbol{A}t_0}\boldsymbol{x}(t_0) + f(\boldsymbol{W_B} + \boldsymbol{W}_C)e^{\boldsymbol{A}^T T}\sum_{i=1}^{N_u}\sum_{j=1}^{N_v}\boldsymbol{H}_{ij}\right]. \tag{56}$$

To obtain the optimal controls, we need to solve for $\boldsymbol{P}_i$ and $\boldsymbol{Q}_j$, which are specified by the following calculations:

$$\boldsymbol{P}_i = \sum_{j=1}^{N_v}\boldsymbol{\lambda}_{ij}(t_0) = \frac{2fe^{\boldsymbol{A}^T(T-t_0)}}{N_u N_v}\sum_{j=1}^{N_v}\left[\boldsymbol{x}_{ij}(T) - \boldsymbol{H}_{ij}\right]$$

$$= \frac{2f}{N_u N_v}e^{\boldsymbol{A}^T(T-t_0)}\left\{\sum_{j=1}^{N_v}\left[e^{\boldsymbol{A}(T-t_0)}\boldsymbol{x}(t_0) - \frac{N_u}{2}e^{\boldsymbol{A}T}\boldsymbol{W_B}e^{\boldsymbol{A}^T t_0}\boldsymbol{P}_i - \frac{N_v}{2}e^{\boldsymbol{A}T}\boldsymbol{W_C}e^{\boldsymbol{A}^T t_0}\boldsymbol{Q}_j\right] - \sum_{j=1}^{N_v}\boldsymbol{H}_{ij}\right\}$$

$$= \frac{2f}{N_u N_v}e^{\boldsymbol{A}^T(T-t_0)}\left[N_v e^{\boldsymbol{A}(T-t_0)}\boldsymbol{x}(t_0) - \frac{N_u N_v}{2}e^{\boldsymbol{A}T}\boldsymbol{W_B}e^{\boldsymbol{A}^T t_0}\boldsymbol{P}_i - \frac{N_v}{2}e^{\boldsymbol{A}T}\boldsymbol{W_C}e^{\boldsymbol{A}^T t_0}\sum_{j=1}^{N_v}\boldsymbol{Q}_j - \sum_{j=1}^{N_v}\boldsymbol{H}_{ij}\right]$$

$$= \frac{2f}{N_u N_v}e^{\boldsymbol{A}^T(T-t_0)}\left[N_v e^{\boldsymbol{A}(T-t_0)}\boldsymbol{x}(t_0) - \frac{N_u N_v}{2}e^{\boldsymbol{A}T}\boldsymbol{W_B}e^{\boldsymbol{A}^T t_0}\boldsymbol{P}_i\right.$$

$$-\frac{f}{N_u}e^{AT}\boldsymbol{W_C}e^{A^TT}\left(\sum_{k=1}^{N_u}\sum_{j=1}^{N_v}\boldsymbol{x}_{kj}(T)-\sum_{k=1}^{N_u}\sum_{j=1}^{N_v}\boldsymbol{H}_{kj}\right)-\sum_{j=1}^{N_v}\boldsymbol{H}_{ij}\Bigg]$$

$$=\frac{2f}{N_uN_v}e^{A^T(T-t_0)}\Bigg[N_ve^{A(T-t_0)}\boldsymbol{x}(t_0)-\frac{N_uN_v}{2}e^{AT}\boldsymbol{W_B}e^{A^Tt_0}\boldsymbol{P}_i$$

$$-\frac{f}{N_u}e^{AT}\boldsymbol{W_C}\left(\left[e^{-(A+A^T)T}+f(\boldsymbol{W_B}+\boldsymbol{W}_C)\right]^{-1}\left[N_uN_ve^{-At_0}\boldsymbol{x}(t_0)-e^{-AT}\sum_{k=1}^{N_u}\sum_{j=1}^{N_v}\boldsymbol{H}_{kj}\right]\right)-\sum_{j=1}^{N_v}\boldsymbol{H}_{ij}\Bigg]. \quad (57)$$

Now the only unknown variable in (57) is $\boldsymbol{P}_i$, and can be solved by

$$\boldsymbol{P}_i=\frac{2f}{N_uN_v}e^{-A^Tt_0}\left(e^{-(A+A^T)T}+f\boldsymbol{W_B}\right)^{-1}\Big[N_ve^{-At_0}\boldsymbol{x}(t_0)$$

$$-\frac{f}{N_u}\boldsymbol{W_C}\left[e^{-(A+A^T)T}+f(\boldsymbol{W_B}+\boldsymbol{W}_C)\right]^{-1}\left(N_uN_ve^{-At_0}\boldsymbol{x}(t_0)-e^{-AT}\sum_{k=1}^{N_u}\sum_{j=1}^{N_v}\boldsymbol{H}_{kj}\right)-e^{-AT}\sum_{j=1}^{N_v}\boldsymbol{H}_{ij}\Bigg]. \quad (58)$$

Similarly, we get that

$$\boldsymbol{Q}_j=\frac{2f}{N_uN_v}e^{-A^Tt_0}\left(e^{-(A+A^T)T}+f\boldsymbol{W_C}\right)^{-1}\Big[N_ue^{-At_0}\boldsymbol{x}(t_0)$$

$$-\frac{f}{N_v}\boldsymbol{W_B}\left[e^{-(A+A^T)T}+f(\boldsymbol{W_B}+\boldsymbol{W_C})\right]^{-1}\left(N_uN_ve^{-At_0}\boldsymbol{x}(t_0)-e^{-AT}\sum_{i=1}^{N_u}\sum_{k=1}^{N_v}\boldsymbol{H}_{ik}\right)-e^{-AT}\sum_{i=1}^{N_u}\boldsymbol{H}_{ij}\Bigg]. \quad (59)$$

Substituting the above two equality into (49) and replacing $t_0$ by $t$ yields the optimal controls in Theorem 3 where $\overline{\boldsymbol{W}}_{\boldsymbol{B}}$ and $\overline{\boldsymbol{W}}_{\boldsymbol{C}}$ are defined in (37). □

**Remark 6.** It is worth pointing out that the target states here can be *arbitrarily* chosen. So the results in Theorem 3 are more generic than those obtained in former sections. Further, if $f$ is large enough, i.e., if $f\overline{\boldsymbol{W}}_{\boldsymbol{B}} >> e^{-(A+A^T)T}$, $f\overline{\boldsymbol{W}}_{\boldsymbol{C}} >> e^{-(A+A^T)T}$, the optimal controls will be specified by:

$$\hat{\boldsymbol{u}}_i=-\boldsymbol{B}^T\boldsymbol{K}(t)\boldsymbol{x}(t)+\boldsymbol{L}_{ui}(t), \quad (60)$$

$$\hat{\boldsymbol{v}}_j=-\boldsymbol{C}^T\boldsymbol{K}(t)\boldsymbol{x}(t)+\boldsymbol{L}_{vj}(t), \quad (61)$$

where

$$\boldsymbol{K}(t) \approx e^{-A^T t} (\overline{\boldsymbol{W}}_{\boldsymbol{B}} + \overline{\boldsymbol{W}}_{\boldsymbol{C}})^{-1} e^{-At} \tag{62}$$

$$\boldsymbol{L}_{\boldsymbol{u}i}(t) \approx -\boldsymbol{B}^T e^{-A^T t} \overline{\boldsymbol{W}}_{\boldsymbol{B}}^{-1} \left[ \frac{\overline{\boldsymbol{W}}_{\boldsymbol{C}} (\overline{\boldsymbol{W}}_{\boldsymbol{B}} + \overline{\boldsymbol{W}}_{\boldsymbol{C}})^{-1} e^{-AT}}{N_u N_v} \sum_{k=1}^{N_u} \sum_{j=1}^{N_v} \boldsymbol{H}_{kj} - \frac{e^{-AT}}{N_v} \sum_{j=1}^{N_v} \boldsymbol{H}_{ij} \right], \tag{63}$$

$$\boldsymbol{L}_{\boldsymbol{v}j}(t) \approx -\boldsymbol{C}^T e^{-A^T t} \overline{\boldsymbol{W}}_{\boldsymbol{C}}^{-1} \left[ \frac{\overline{\boldsymbol{W}}_{\boldsymbol{B}} (\overline{\boldsymbol{W}}_{\boldsymbol{B}} + \overline{\boldsymbol{W}}_{\boldsymbol{C}})^{-1} e^{-AT}}{N_u N_v} \sum_{i=1}^{N_u} \sum_{k=1}^{N_v} \boldsymbol{H}_{ik} - \frac{e^{-AT}}{N_u} \sum_{i=1}^{N_u} \boldsymbol{H}_{ij} \right]. \tag{64}$$

If the target states satisfy the constraint in (9), namely, $\boldsymbol{H}_{ij} - \boldsymbol{H}_{i'j} = \boldsymbol{H}_{ij'} - \boldsymbol{H}_{i'j'}$ and if $i' = j' = 1$, then (63) and (64) reduce to

$$\boldsymbol{L}_{\boldsymbol{u}i}(t) \approx \boldsymbol{B}^T e^{-A^T t} \left[ \overline{\boldsymbol{W}}_{\boldsymbol{B}}^{-1} e^{-AT} \boldsymbol{H}_{i1} + (\boldsymbol{I} + \overline{\boldsymbol{W}}_{\boldsymbol{C}}^{-1} \overline{\boldsymbol{W}}_{\boldsymbol{B}})^{-1} \left( \overline{\boldsymbol{W}}_{\boldsymbol{C}}^{-1} e^{-AT} \sum_{j=1}^{N_v} \frac{\boldsymbol{H}_{1j} - \boldsymbol{H}_{11}}{N_v} - \overline{\boldsymbol{W}}_{\boldsymbol{B}}^{-1} e^{-AT} \sum_{k=1}^{N_u} \frac{\boldsymbol{H}_{k1}}{N_u} \right) \right], \tag{65}$$

$$\boldsymbol{L}_{\boldsymbol{v}j}(t) \approx \boldsymbol{C}^T e^{-A^T t} \left[ \overline{\boldsymbol{W}}_{\boldsymbol{C}}^{-1} e^{-AT} \boldsymbol{H}_{1j} + (\boldsymbol{I} + \overline{\boldsymbol{W}}_{\boldsymbol{B}}^{-1} \overline{\boldsymbol{W}}_{\boldsymbol{C}})^{-1} \left( \overline{\boldsymbol{W}}_{\boldsymbol{B}}^{-1} e^{-AT} \sum_{i=1}^{N_u} \frac{\boldsymbol{H}_{i1} - \boldsymbol{H}_{11}}{N_u} - \overline{\boldsymbol{W}}_{\boldsymbol{C}}^{-1} e^{-AT} \sum_{k=1}^{N_v} \frac{\boldsymbol{H}_{1k}}{N_v} \right) \right]. \tag{66}$$

Note that the right hand side of (62), (65) and (66) are exactly the same as (35)-(36) in Theorem 2.

**Remark 7.** Under the feedback controls (41) and (42), the system evolves along the optimal trajectory $\boldsymbol{x}_{ij}(t)$ with dynamics:

$$\dot{\boldsymbol{x}}_{ij}(t) = [\boldsymbol{A} - \boldsymbol{B}\boldsymbol{K}_{\boldsymbol{u}}(t) - \boldsymbol{C}\boldsymbol{K}_{\boldsymbol{v}}(t)] \boldsymbol{x}_{ij}(t) + \boldsymbol{B}\boldsymbol{L}_{\boldsymbol{u}i}(t) + \boldsymbol{C}\boldsymbol{L}_{\boldsymbol{v}j}(t).$$

When $f$ is large enough, the final state of the system at the terminal time, $T$, can be represented by

$$\boldsymbol{x}_{ij}(T) \approx \frac{1}{N_u} \sum_{i=1}^{N_u} \boldsymbol{H}_{ij} + \frac{1}{N_v} \sum_{j=1}^{N_v} \boldsymbol{H}_{ij} - \frac{1}{N_u N_v} \sum_{i=1}^{N_u} \sum_{j=1}^{N_v} \boldsymbol{H}_{ij}. \tag{67}$$

The collection of final states correlate with the set of target states in the following way

$$\sum_{i=1}^{N_u} \sum_{j=1}^{N_v} \boldsymbol{x}_{ij}(T) \approx \sum_{i=1}^{N_u} \sum_{j=1}^{N_v} \boldsymbol{H}_{ij}.$$

This point can also be observed from (56) directly. Again, if $\boldsymbol{H}_{ij} - \boldsymbol{H}_{i'j} = \boldsymbol{H}_{ij'} - \boldsymbol{H}_{i'j'}$ is satisfied, (67) yields the expected result $\boldsymbol{x}_{ij}(T) \approx \boldsymbol{H}_{ij}$.

## VI. Conclusions

In the present paper we have investigated the design problem of affine choice-based action systems in which multiple agents cooperatively apply controls based on their independent choices. A systematic design methodology for optimal *target-achieving* control of general multi-agent affine systems has been established. In the meanwhile, we have derived a set of optimal *target-approaching* controls, which can be seen as a general solution for targeting arbitrary states in finite time with minimum average control cost. It is shown in these results that the average control cost is closely related to the cardinality of each set of control choices, number of agents, distances between the given target states and the initial state, and the level of asymmetry of the given terminal states in their spatial distribution.

It is worth mentioning that the solution to the stated problem is based on the assumption that the system is controllable by each agent individually. If the system under investigation is not individually controllable but jointly controllable, the problem remains unsolved. What seems apparent in this case is that some level of communication on the choice information would need to be made by the agents. The framework described in [23], [25] and [26] then become relevant.

## References

[1]. M. Athans and P. L. Falb, *Optimal control: an introduction to the theory and its applications*. McGraw-Hill, New York, U.S.A., 1966.

[2]. J. Baillieul and W. S. Wong, “The standard part problem and the complexity of control communication,” in *Proc. IEEE Conference on Decision and Control,* Shanghai, 2009.

[3]. L. Barcelos-Oliveira and E. Camponogara,“Multi-agent model predictive control of signaling split inn urban traffic networks,” *Transportation Research Part C,* vol. 18, pp. 120-139, 2010.

[4]. D. Bauso, L. Giarre, and R. Pesenti, “Mechanism design for optimal consensus problems,” in *Proc. IEEE Conference on Decision and Control*, pp. 3381-3386, 2006.

[5]. R. W. Brockett, “Minimizing attention in a motion control context,” in *Proc. IEEE Conference on Decision and Control*, pp. 3349-3352, 2003.

[6]. R. W. Brockett, “Optimal control of the Liouville Equation,” in Proc. *International Conference on Complex Geometry and Related Fields*, pp. 23-35, 2007.

[7]. J. A. Fax, “Optimal and cooperative control of vehicle formations,” Ph.D. thesis. California Institute of Technology, 2002.

[8]. L. Faybusovich, T. Mouktonglang, “Multi-target linear-quadratic control problem and second-order cone programming,” *Systems and Control Letters*, vol. 52, pp. 17-23, 2004.

[9]. G. Inalhan, D. M. Stipanovic, and C. J. Tomlin, “Decentralized optimization, with application to multiple aircraft coordination,” in *Proc. conference on decision and control*, pp. 1147-1155, 2002.

[10]. F. Kunwar, B. Benhabib, “Rendezvous-guidance trajectory planning for robotic dynamic obstacle avoidance and interception,” *IEEE Transactions on Systems, Man, and Cybernetics - Part B: Cybernetics*, vol. 36, no. 6, pp. 1432-1441, 2006.

[11]. T. Li, and J. F. Zhang, “Decentralized tracking-type games for multi-agent systems with coupled ARX models: asymptotic Nash equilibria,” *Automatica*, vol. 44, pp. 713-725, 2008.

[12]. W. Li, “Stability analysis of swarms with general topology,” *IEEE Transactions on Systems, Man, and Cybernetics - Part B: Cybernetics*, vol. 38, no.4, pp. 1084-1097, 2008.

[13]. H. Liu, and J. Li, “Terminal sliding mode control for spacecraft formation flying,” *IEEE Transactions on Aerospace and Electronic Systems*, vol. 45, no. 3, pp. 835-846, 2009.

[14]. J. S. Meditch, “Synthesis of a class of linear feedback minimum energy controls,” *IEEE Transactions on Automatic Control*, 8(4): 376-378, 1963.

[15]. F. Michaud, P. Lepage, P. Frenette, D. Letourneau, N. Gaubert, “Coordinated maneuvering of automated vehicles in platoons,” *IEEE Transactions on Intelligent Transportation Systems*, vol. 7, no.4, pp. 437-447, 2006.

[16]. R. Olfati-Saber, “Flocking for multi-agent dynamic systems: algorithms and theory,” *IEEE*

*Transactions on Automatic Control*, vol. 51, no. 3, pp. 401-420, 2006.

[17]. R. L. Raffard, C. J. Tomlin, and S. P. Boyd, "Distributed optimization for cooperative agents: application to formation flight," in *Proc. conference on decision and control*, pp. 2453-2459, 2004.

[18]. W. Ren, "Consensus tracking under directed interaction topologies: algorithms and experiments," *IEEE Transactions on Control Systems Technology*, vol. 18, no. 17, pp. 230-237, 2010.

[19]. E. Semsar-Kazerooni and K. Khorasani, "Optimal control and game theoretic approaches to cooperative control of a team of multi-vehicle unmanned systems," in *Proc. IEEE international conference on networking, sensing and control*, pp. 628-633, 2007.

[20]. E. Semsar-Kazerooni and K. Khorasani, "Optimal consensus algorithms for cooperative teams of agents subject to partial information," *Automatica*, vol. 44, pp. 2766-2777, 2008.

[21]. E. Semsar-Kazerooni and K. Khorasani, "Multi-agent team cooperation: a game theory approach," *Automatica*, vol. 45, pp. 2205-2213, 2009.

[22]. Y. Shoham and K. Leyton-Brown, *Multiagent Systems: Algorithmic, Game-Theoretic and Logical Foundations*. Cambridge University Press, 2009.

[23]. W. Wojsznis, A. Mehta, P. Wojsznis, D. Thiele, T. Blevins, "Multi-objective optimization for model predictive control," *ISA Transactions*, vol. 46, pp. 351-361, 2007.

[24]. W. S. Wong, "Control communication complexity of distributed control systems," *SIAM Journal of Optimization and Control*, vol. 48 no.3, pp. 1722-1742, 2009.

[25]. W. S. Wong and J. Baillieul, "Control communication complexity of nonlinear systems," *Communications in Information and Systems*, vol. 9, no.1, pp. 103-140, 2009.

[26]. W. S. Wong and J. Baillieul, "Control communication complexity of distributed actions," to appear in the *IEEE Transactions on Automatic Control*.

[27]. A. C. C. Yao, "Some complexity questions related to distributive computing," in *Proc. Annual ACM Symposium on Theory of Computing*, 1979.